\documentclass[letterpaper]{article}
\usepackage{iccc}

\usepackage{times}
\usepackage{helvet}
\usepackage{courier}

\usepackage{xcolor}
\usepackage{subfig}
\usepackage{amsmath} 
\usepackage{hyperref}  
\usepackage{url}            
\usepackage{natbib} 

\usepackage{float}
\usepackage{graphicx}

\definecolor{ddarkbrown}{rgb}{0.5,0.2,0.05} \definecolor{bbluegray}{rgb}{0.05,0,0.5}

\newcommand{\BEAS}{\begin{eqnarray*}}
\newcommand{\EEAS}{\end{eqnarray*}}
\newcommand{\BEA}{\begin{eqnarray}}
\newcommand{\EEA}{\end{eqnarray}}
\newcommand{\BEQ}{\begin{equation}}
\newcommand{\EEQ}{\end{equation}}
\newcommand{\BIT}{\begin{itemize}}
\newcommand{\EIT}{\end{itemize}}
\newcommand{\BNUM}{\begin{enumerate}}
\newcommand{\ENUM}{\end{enumerate}}

\newcommand{\BA}{\begin{array}}
\newcommand{\EA}{\end{array}}

\title{Interactive Neural Style Transfer with Artists}
\author{
   Thomas Kerdreux\thanks{Equal contribution.}\\
   Department of Computer Science\\
   ENS, PSL University \& INRIA\\
   Paris, France\\
   \And
   Louis Thiry\footnotemark[1]\\
   Department of Computer Science\\
   ENS, CNRS, PSL University\\
   Paris, France\\
   \And
   Erwan Kerdreux\\
   Departement of Design\\
   ENS Paris-Saclay\\
   Palaiseau, France\\
 }

\begin{document} 
\maketitle
\begin{abstract}
We present interactive painting processes in which a painter and various neural style transfer algorithms interact on a real canvas.
Understanding what these algorithms' outputs achieve is then paramount to describe the creative agency in our interactive experiments.
We gather a set of paired painting-pictures images and present a new evaluation methodology based on the \textit{predictivity} of neural style transfer algorithms.
We point some algorithms' instabilities and show that they can be used to enlarge the diversity and pleasing oddity of the images synthesized by the numerous existing neural style transfer algorithms.
This diversity of images was perceived as a source of inspiration for human painters, portraying the machine as a \textit{computational catalyst}.
\end{abstract}

\section{Introduction}

{\em Neural style transfer} \citep{gatys2015neural}, which seeks at rendering the content of one image using the  style of another, provides impressive results as it takes advantage of the rich hierarchical representation of images produced by convolutional neural networks (CNN) to quantify the style and content of images.
The many ways to manipulate these complex maps, as well as their increasing ease of implementation, have hence underpinned the development of a plethora of successful methods in this area of computational artistic rendering.

Several evaluation techniques have been developed to compare all different methods.
On the one hand, many methods focus on quantifying how much a neural style transfer method attains a numerical objective.
These are good engineering indicators, but we highlight that they are not necessarily relevant at measuring the quality of the outputs of style transfer algorithm.
On the other hand, qualitative evaluation methods typically consist of collecting a large number of subjective impressions on the algorithms' outputs.
It provides average scores on the content preservation and style quality of the algorithms' outputs but does not reveal the specificity of an algorithm compare to another. 

To have a more precise characterization of these algorithms, we introduce a new evaluation methodology based on the \textit{predictivity} of neural style transfer algorithms and gather a set of paired painting-photographic for this evaluation.
The \textit{predictivity} consists of assessing whether or not the algorithms' outputs are close to an existing painting when using this painting as the style image and the associated photograph as the content image.
This is also a crucial point in the computational creativity perspective as some outputs are deemed interesting while bearing not much resemblance with the initial painting, \textit{i.e.} with what the painter did.

In addition, when showing to artists some outputs of style transfer algorithms using their own paintings as style images, they often do not recover their practice. However, they sometimes identify inspiring aspects
in the various outputs of different algorithms, implicitly acknowledging their computational creativity.
This naturally led us to painting processes with artists, who could not only edit groups of style transfer outputs, but use them as basic elements to widen their style.
This constructively interlaces the agency attribution of the algorithms part in the creative process.

We further encouraged this complexity by exploring these algorithms in the real world, where the outputs are projected onto a real canvas, the classical space for human painters.
Human and machine contributions are then mingled in a single canvas. Interestingly then, to help the observer that seeks to untangle the contribution of each agent, the canvas can be shown together with the various computational suggestions.
In such creative processes, the algorithms were experienced as \textit{computational catalysts} to human creativity, a middle ground between agents that are creative on their own and still technical tools.

\paragraph{Outline.}
We first describe the new methodology for qualitative evaluation of different style transfer algorithms.
We then question the relevance of existing quantitative evaluations giving a simple example where improving the quantitative criterion of an algorithm does not improve the quality of the stylization quality of the outputs.
We then indicate that some approaches to neural style transfer do not satisfy a basic property, which leads to an instability behaviour that ultimately allows reinforcing the diversity of style transfer outputs.

Based on all these observations, we present various interactive painting experiments between human and style transfer outputs.
This leads us to the notion of \textit{computational catalysts} that help to characterize the algorithms' contribution in our specific settings.

\begin{figure}[H]
    \centering
    \subfloat[content]{{\includegraphics[width=0.275\linewidth]{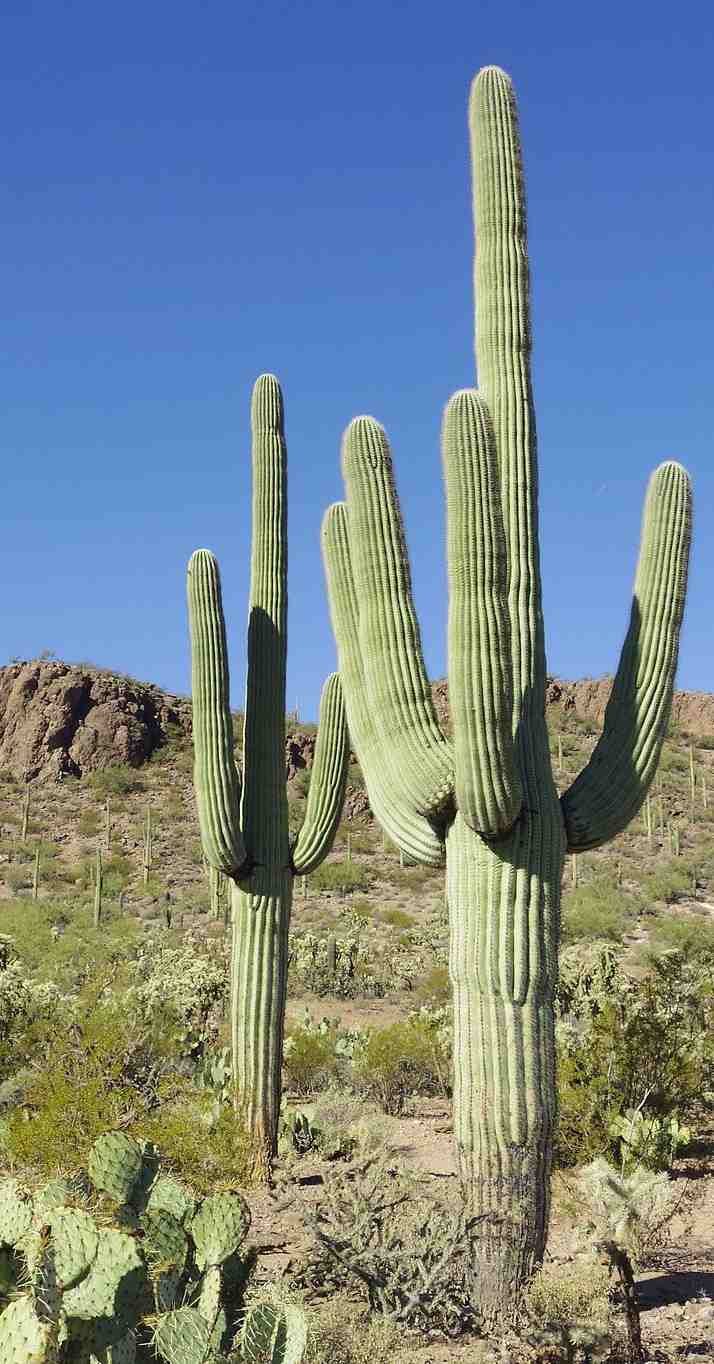} }}%
    \subfloat[style]{{\includegraphics[width=0.525\linewidth]{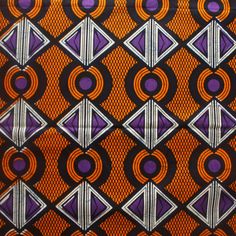} }}\\
    \hfill
    \subfloat[AdaIn]{{\includegraphics[width=0.195\linewidth]{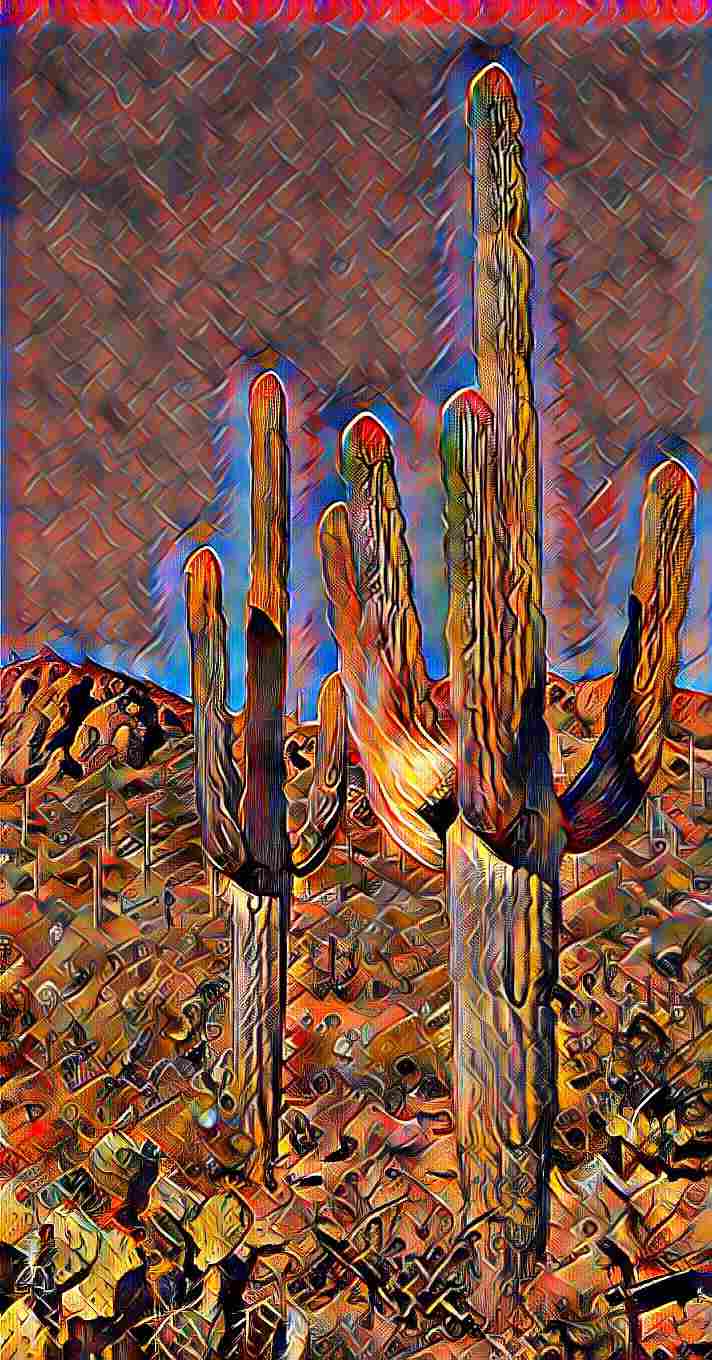} }}%
    \subfloat[MST]{{\includegraphics[width=0.195\linewidth]{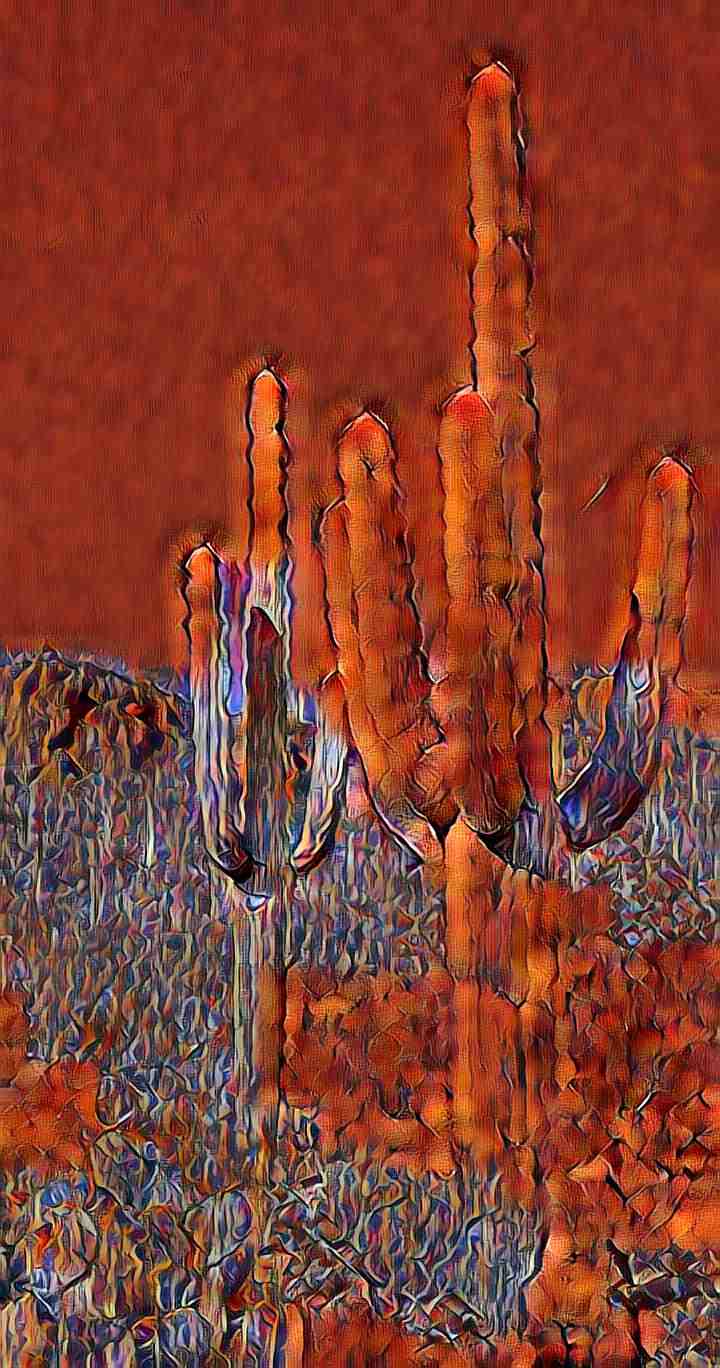} }}%
    \subfloat[WCT]{{\includegraphics[width=0.195\linewidth]{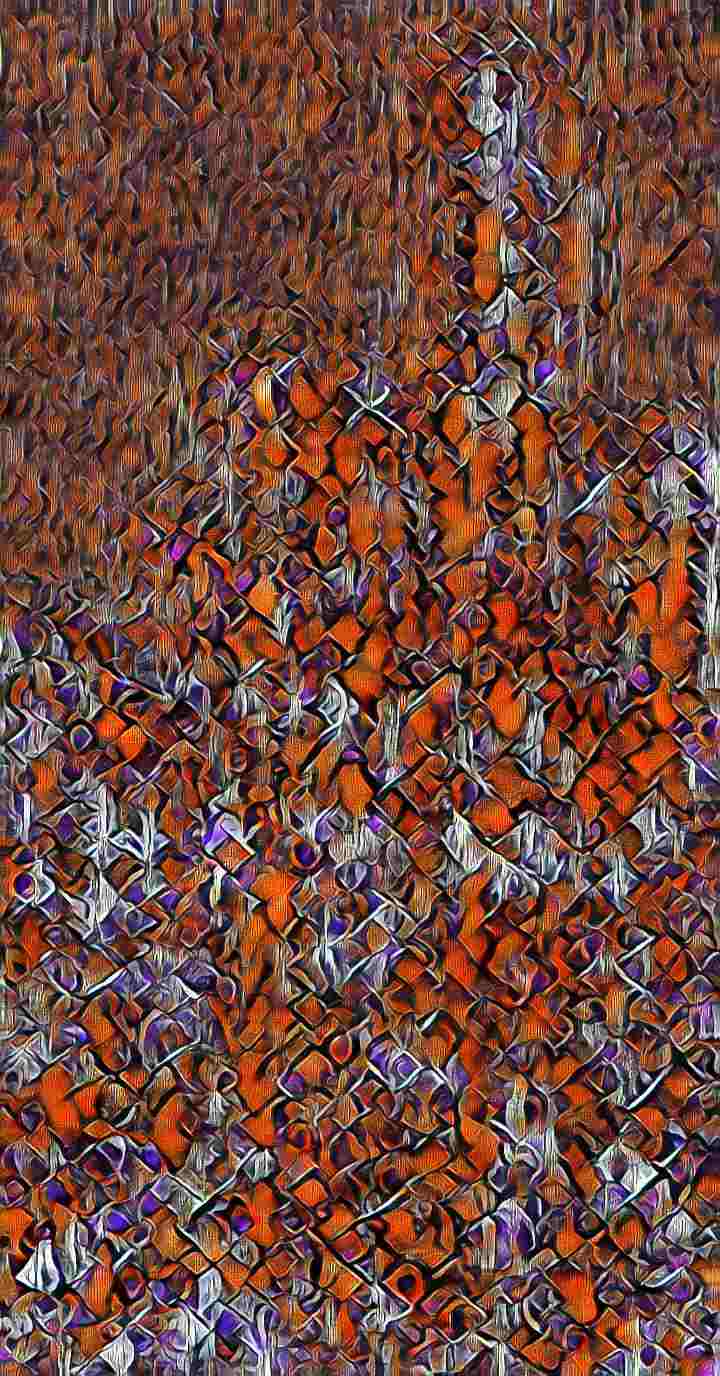} }}%
    \subfloat[STROTSS]{{\includegraphics[width=0.195\linewidth]{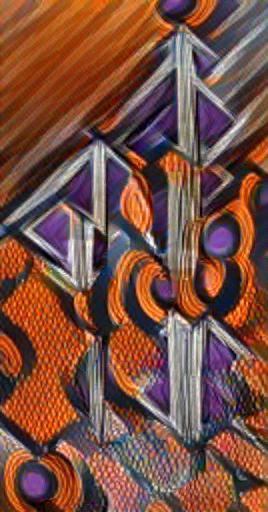} }}%
    \subfloat[Gatys]{{\includegraphics[width=0.195\linewidth]{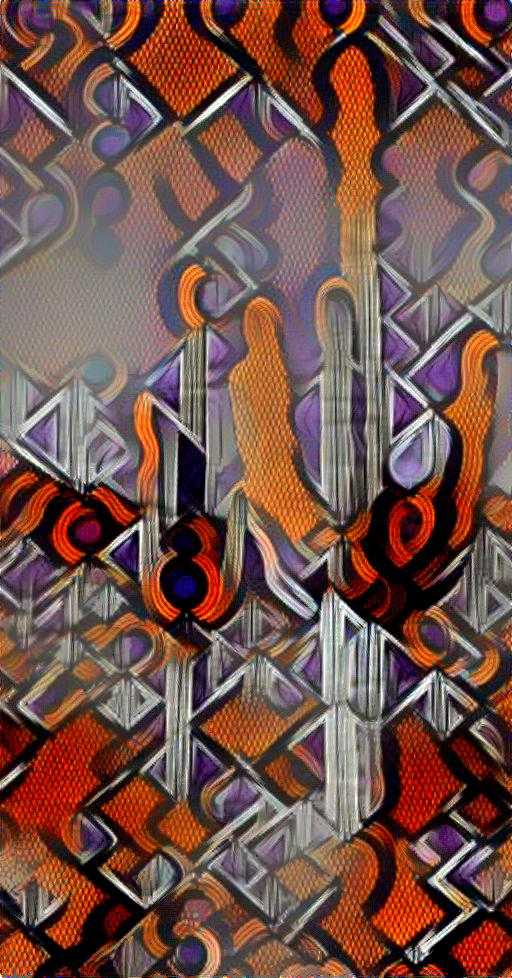} }}\\
    \caption{Outputs of neural style transfer algorithms AdaIn \citep{huang2017arbitrary}, MST \citep{zhang2019multimodal}, WCT \citep{li2017universal}, STROTSS \citep{TransferTransport} and Gatys \citep{gatys2015neural} on the same \{content, style\} pair.}
    \label{fig:style_transfer_ex}
\end{figure}

\section{Evaluating Neural Style Transfer Methods}
Neural methods for style transfer originated with the optimization-based technique of \cite{gatys2015neural} which leverages the image features extracted from convolutional neural networks (CNN). To speed up the process as well as to have access to representations of a particular painting style, \citep{li2016precomputed,ulyanov2016texture,johnson2016perceptual} then proposed to train a neural network dedicated to a particular style, enabling to do neural style transfer of an image with a single forward pass instead of a full optimization procedure.
Later on, universal neural style transfer methods were developed to transfer any kind of style to a content image, again with a single forward pass \citep{ghiasi2017exploring,li2017universal,huang2017arbitrary}.
These approaches are much faster than the optimization-based approaches but they suffer from the well-documented instabilities of neural network \citep{szegedy2013intriguing}.
We show that a specific instability that, to the best of our knowledge, has not been pointed out yet, can notably be beneficial as it enlarges the creative possibilities of neural style transfer.

Alternatively, to explore other creative opportunities of such algorithms,
several user control methods have been developed using for example semantic correspondences \citep{lu2017decoder,gatys2017controlling,TransferTransport}, 
allowing to hand-tune the colour histograms  \citep{gatys2017controlling} or the scale of the patterns \citep{risser2017stable}. It is also possible to transfer multiple styles \citep{WassersteinStyleTransfer, cheng2019structure} at once.

Many works are still exploring different neural style transfer approaches, for instance working with histogram losses \citep{risser2017stable}, using various relaxation of optimal transport \citep{TransferTransport,WassersteinStyleTransfer,kotovenko2019content} or trying to match semantic patterns in content and style images \citep{GraphCut19}.
All these methods achieve impressive plastic results, but they are hard to characterize one w.r.t. the other. They may not yet actually stylize an image in the many ways a human would. We thus study the question of evaluation methods for style transfer.

\subsection{Qualitative evaluation}
\begin{figure}[ht]
    \centering
    \subfloat[Hassam painting]{{\includegraphics[width=0.32\linewidth]{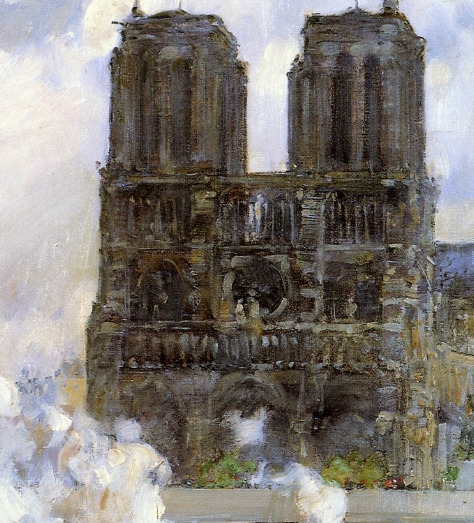} }}
    \subfloat[photograph]{{\includegraphics[width=0.32\linewidth]{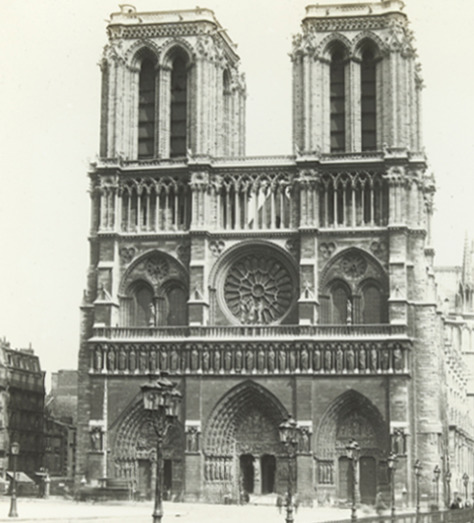} }}
     \subfloat[semantic mask]{{\includegraphics[width=0.32\linewidth]{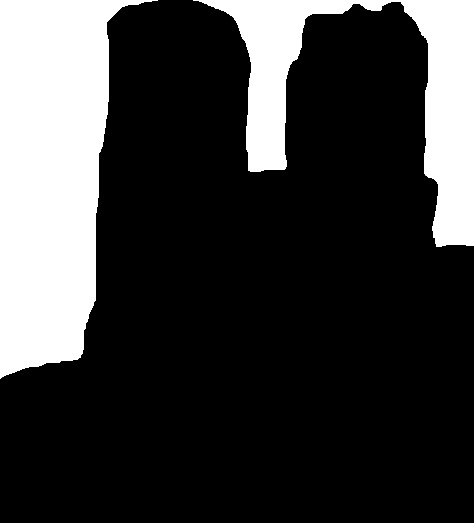} }}\\
    \hfill
    \centering
    \subfloat[MST]{{\includegraphics[width=0.32\linewidth]{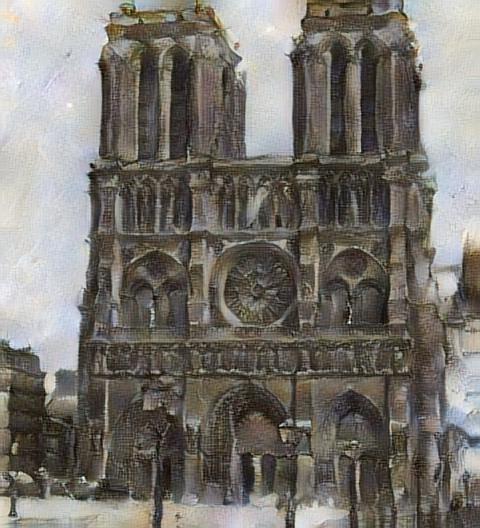} }}%
    \subfloat[STROTSS]{{\includegraphics[width=0.32\linewidth]{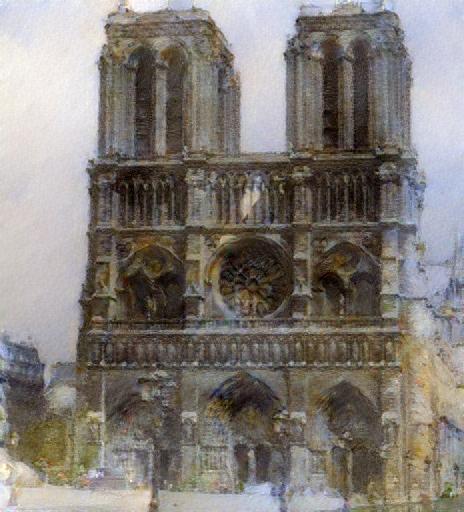} }}%
    \subfloat[Gatys]{{\includegraphics[width=0.32\linewidth]{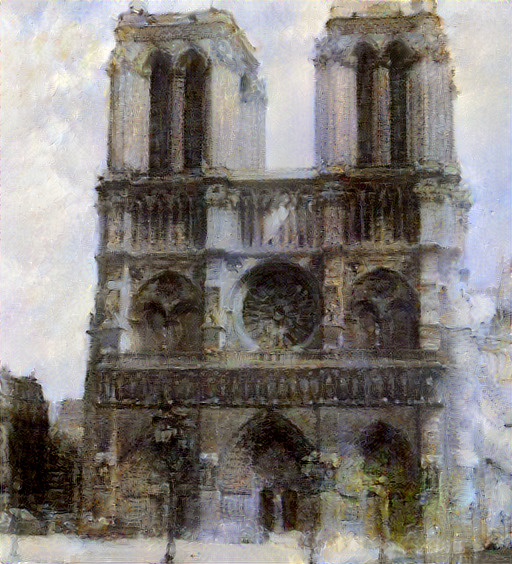} }}\\
    \hfill
    \caption{Detail of \textit{Quai St Michel}, Childe Hassam, corresponding photo, semantic mask and style transfer outputs.}
    \label{fig:style_transfer_pair_paris}
\end{figure}

A natural way of evaluating a neural style transfer method is to measure the content preservation and the stylization quality of the outputs.
The variety of possible input images for content and style makes this task difficult in general.
For example Gatys succeeds in transferring the style of Van Gogh's \textit{Starry night} but the examples shown in figures \ref{fig:style_transfer_ex} and \ref{fig:style_transfer_pair_paris} show notable artefacts.
Such an evaluation can still be done gathering a large number of responses as \cite{TransferTransport} did to measure the content or style preservation of their method compared to the others.
Results showed that their method (STROTSS) offered on average the best trade-off between content and style preservation but does not say in what sense the style and content are better preserved.

To have a systematic and more refined comparison, we propose to study the \textit{predictivity} of style transfer algorithm: does an algorithm stylize the image in a way a painter would have done? 
Precisely, when considering a photograph as a content image and a figurative painting of this image as a style
image, one can compare the output of the neural style transfer algorithm with the figurative painting and further judge whether the style transfer technique succeeds in predicting the painting, and if not, try to characterize how it differs from it.

\begin{figure}[ht]
    \centering
    \subfloat[Monet painting]{{\includegraphics[width=0.32\linewidth]{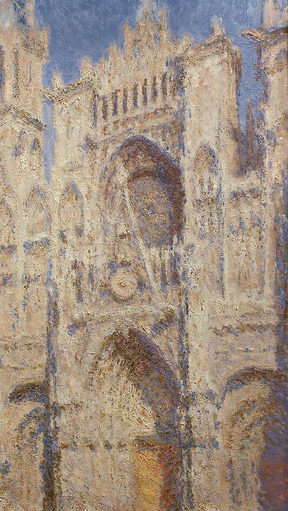} }}
    \subfloat[photograph]{{\includegraphics[width=0.32\linewidth]{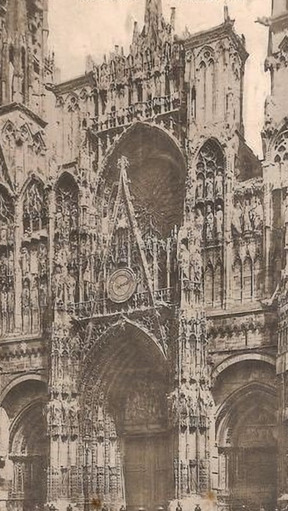} }}%
    \subfloat[semantic mask]{{\includegraphics[width=0.32\linewidth]{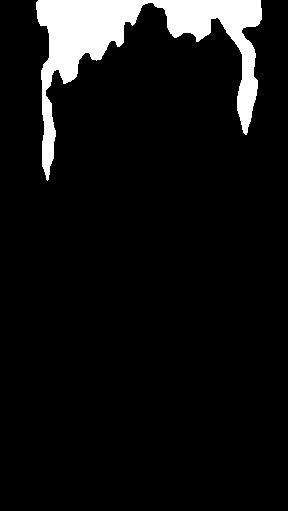} }}\\
    \hfill
    \centering
    \subfloat[WCT]{{\includegraphics[width=0.32\linewidth]{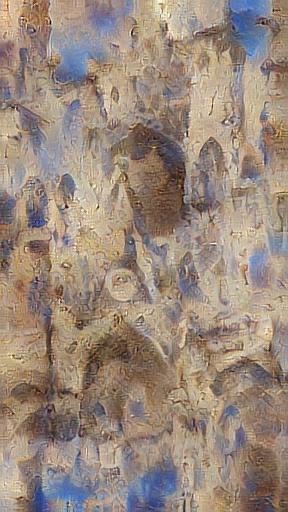} }}%
    \subfloat[STROTSS]{{\includegraphics[width=0.32\linewidth]{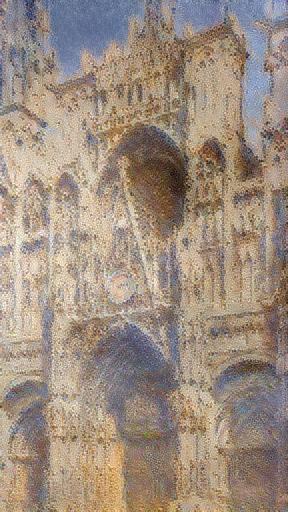} }}%
    \subfloat[Gatys]{{\includegraphics[width=0.32\linewidth]{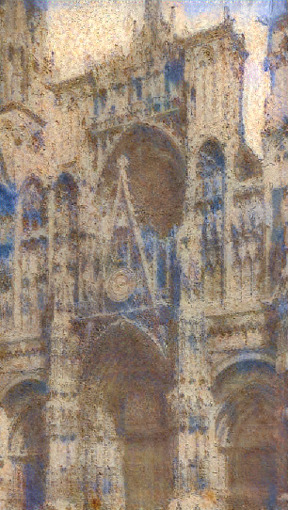} }}\\
    \hfill
    \caption{Detail of \textit{Le Portail de la cathédrale de Rouen au soleil}, Monet, corresponding photograph and semantic mask and style transfer outputs}
    \label{fig:style_transfer_pair_rouen}
\end{figure}

Such pairs of photograph and content-preserving paintings are not readily available; landscapes are constantly changing, face portraits are rarely faithful to the original as well as we rarely possess the photograph of the model. Building paintings, however, is a good class of paintings for the proposed study.
We thus construct a set of photographic-painting pairs, see Figures \ref{fig:style_transfer_pair_paris}-\ref{fig:style_transfer_pair_rouen} for instance, focusing on two famous buildings that have inspired numerous artists: the \textit{Notre Dame de Paris Cathedral} and the \textit{Notre Dame de Rouen Cathedral}.
We gathered pictures of paintings of a three-quarter view of the facade of \textit{Notre Dame de Paris Cathedral} by Utrillo,  Matisse,  Luce, Marquet, Barthold, Guillaumin, and Hassam.
Particularly interesting for our study, Claude Monet made a series of about forty paintings capturing the facade of \textit{Notre Dame de Rouen Cathedral} from nearly the same viewpoint at different times of the day and year and under different meteorological and lighting conditions \cite[p.~656]{livrecathedralerouen}.
As some methods can use semantic masks to specify corresponding regions of the content and style image, we add a semantic mask to each pair.

With this set, qualitative evaluation can be done more systematically and less arbitrarily; in the example shown in Figure \ref{fig:style_transfer_pair_rouen}, STROTSS output is qualitatively the closest to the Monet painting, especially for the lightening effect on the door and the left of the portal.
Gatys and WCT suffer from spatial inconsistency as the blue sky is replaced by a sunlight halo in the first one and the background is hardly distinguishable in the second one.
We release this set together with the outputs of the style transfer algorithms to facilitate and systematize the qualitative evaluation of neural style transfer techniques.

\subsection{Quantitative evaluation}

Numerical evaluation methods have the benefit of being more systematic and objective.
However, we point here that most neural style transfer evaluation methods are specific to certain algorithms and are not always relevant for the stylization quality of the output.

In computer vision, perceptual losses are becoming the standard to compare the visual similarity between images \citep{zhang2018unreasonable}.
These methods based on CNN features comparison offer state of the art performance on image similarity judgment datasets using different CNN architectures.
Since neural style transfer originally consists of optimizing an image in order to match the CNN features of another style image, the perceptual loss between the outputs and the target style image might be artificially small despite notable perceptual differences.

Other numerical evaluation techniques were proposed; \cite{sanakoyeu2018style} test whether a  pre-trained neural network for artist classification on real paintings succeeds in classifying the artist of the style image based on an algorithm’s output.
\cite{jing17review} consider comparing saliency maps between images since the spatial integrity and coherence of the saliency maps should remain similar after style transfer. 
Moreover, as neural style transfer relies on a certain quantification of the style based on CNN features, \cite{jing17review} propose to evaluate how much the optimization objective is achieved in style transfer.
We show in the following case that improving the optimization objective is not necessarily related to the visual quality of the output.

Optimization-based neural style transfer methods consist in optimizing the pixels of an image $I$ to minimize a loss $l$.
This loss $l$ is usually the sum of a content loss $l_c(I, I_c)$ measuring the content similarity between $I$ and the content image $I_c$ and a style loss $l_s(I, I_s)$ measuring the style similarity between $I$ and the style image $I_s$.
In the STROTSS method, \cite{TransferTransport} define the style loss as the Earth Movers Distance (EMD) between CNN features of the image $I$ and the style image $I_s$.
Given the CNN features $\Phi(I), \Phi(I_s)$ of the images $I, I_s$, we compute the distance matrix $C^{I, I_s}$ between $\Phi(I)$ and $\Phi(I_s)$ and the EMD is defined as the solution of the following optimization problem
\[
    \text{EMD} (I, I_s) =  \min_{\substack{T \geq 0 \\ \sum_j T_{ij} = 1 / m\\\sum_i T_{ij} = 1 / n}}  \sum_{ij} T_{ij} C^{I, I_s}_{ij} \ .
\] 

Exact EMD computations are too expensive for neural style transfer applications, and a relaxed EMD (REMD) is used in STROTSS.
It consists in taking the maximum of two simple lower bounds of the EMD, each obtained removing one of the two linear constraints sets $\sum_{i\text{ or }j} T_{ij}$ applied on the transport plan $T$
\begin{align*}
    \text{REMD} (I, I_s) =
        \text{max } \left(
             \begin{array}{c}
                \underset{T \geq 0, \  \sum_{\mathbf{i}} T_{ij} = 1 / n}{\text{min }} \  \underset{ij}{\sum} \  T_{ij} C^{I, I_s}_{ij} \ ,\\
                \underset{T \geq 0, \  \sum_{\mathbf{j}} T_{ij} = 1 / m}{\text{min }} \  \underset{ij}{\sum} \ T_{ij} C^{I, I_s}_{ij}
             \end{array}
        \right)~.
\end{align*}

Despite the use of this loose relaxation, the human evaluation done via Amazon Mechanical Turk (AMT) indicates that STROTSS statistically offers the best style/content trade-off compared to the other neural style transfer techniques \citep[\S 4]{TransferTransport} in the opinion of the AMT workers. Experiments done with artists confirmed this trend as the artists were mostly impressed by results produced by STROTSS. 

The authors mentioned that a better approximation may yield better style transfer results.
Sinkhorn-distance \citep{cuturi2013sinkhorn} in its log-domain stabilized version \citep{schmitzer2019stabilized} is a good candidate to this purpose.
We thus replaced the relaxed earth movers distance REMD by the Sinkhorn earth movers distance
\[
    \text{SEMD}_{\epsilon} (I, I_s) =  \min_{\substack{T \geq 0 \\ \sum_j T_{ij} = 1 / m\\\sum_i T_{ij} = 1 / n}}  \sum_{ij} T_{ij} C^{I, I_s}_{ij} - \epsilon h(T)
\]
where $h$ is the entropy of the transport plan $T$
\[ h(T) = -\sum_{ij} T_{ij}\log T_{ij} \]
and $\epsilon$ is the entropic regularization parameter.
The corresponding optimization problem is convex and is solved iteratively with a fixed number of iterations $N$.
$\text{SEMD}_{\epsilon}$ is an upper-bound of the EMD and it converges to the exact EMD as $\epsilon$ goes to $0$.
We release a Pytorch \citep{paszke2019pytorch} implementation of STROTSS including the SEMD.

\begin{figure}[h]
\includegraphics[width=0.455\linewidth]{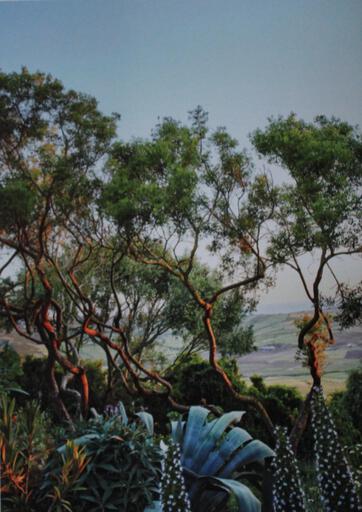}
\includegraphics[width=0.455\linewidth]{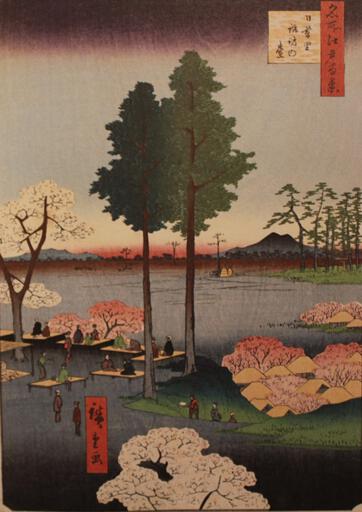}\\
\includegraphics[width=0.3\linewidth]{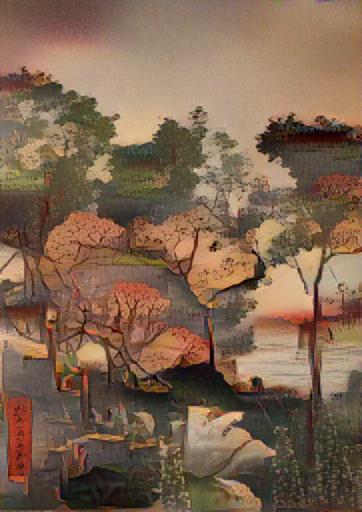}
\includegraphics[width=0.3\linewidth]{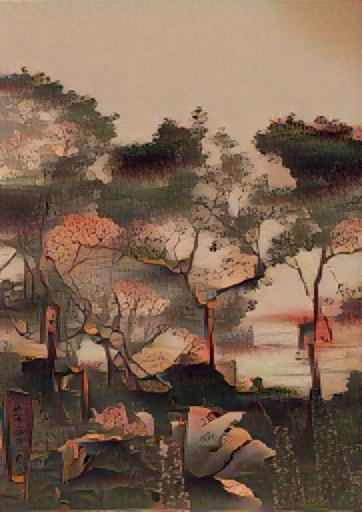}
\includegraphics[width=0.3\linewidth]{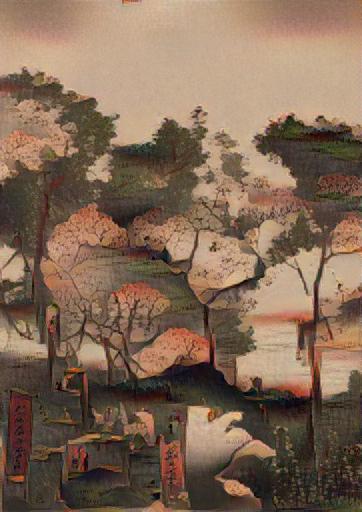}
\caption{Left to right, (top) content, style,  (bottom) outputs using REMD, SEMD $\epsilon = 1e^{-1}, N = 30$ and SEMD $\epsilon = 1e^{-3}, N = 50$.}
\label{fig:STROTSScomparision}
\end{figure}

Figure \ref{fig:STROTSScomparision} shows a comparison of experimental results, suggesting that getting much closer to the mathematical quantification of the style does not necessarily lead to more relevant results, and numerical evaluation of how much the mathematical objective is achieved is not essential from a visual perspective.

In the same idea, the instability phenomena that are commonly assumed to be detrimental in the neural networks literature (e.g. adversarial examples), can qualitatively increase the creative possibilities of neural style transfer.

\subsection{Instability phenomena}
\label{ssec:instability_style_transfert}

Neural style transfer instabilities have been pointed out by \cite{risser2017stable} and \cite{gupta2017characterizing} in the case of real-time style transfer for videos.
The aim is to identify and remove the time-inconsistent artefacts that create unpleasing effects.
Here we outline instabilities stemming from another type of inconsistency and propose to take advantage of them.

\begin{figure}
    \centering
    \includegraphics[width=0.24\linewidth]{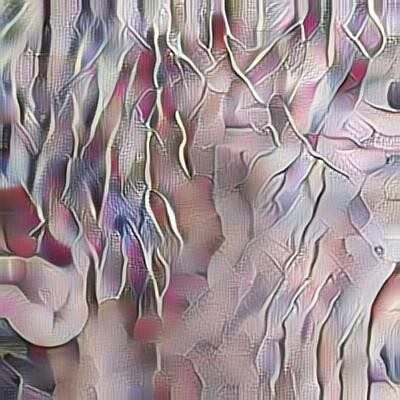}
    \includegraphics[width=0.24\linewidth]{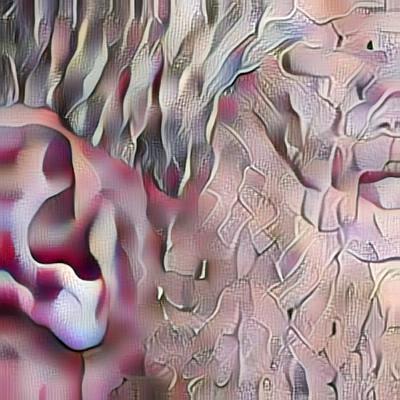}
    \includegraphics[width=0.24\linewidth]{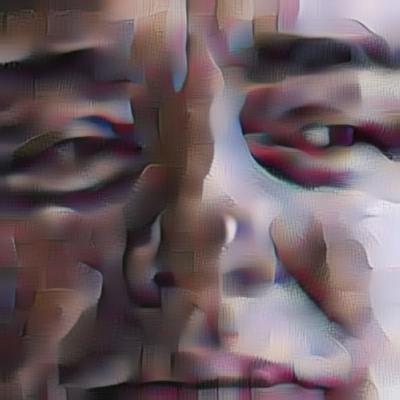}
    \includegraphics[width=0.24\linewidth]{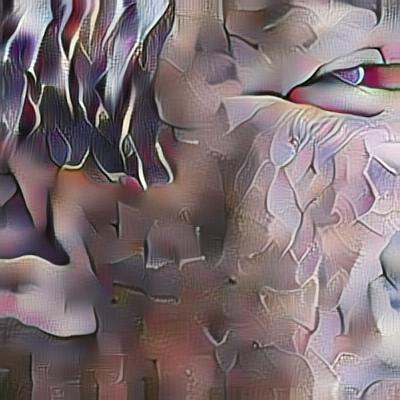}
    \includegraphics[width=0.24\linewidth]{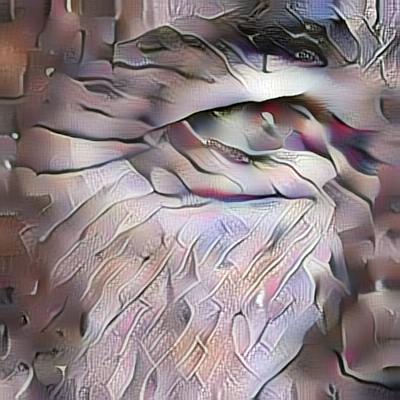}
    \includegraphics[width=0.24\linewidth]{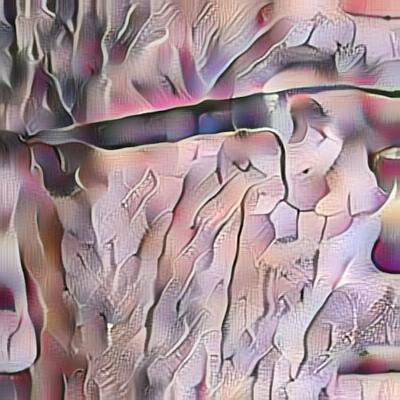}
    \includegraphics[width=0.24\linewidth]{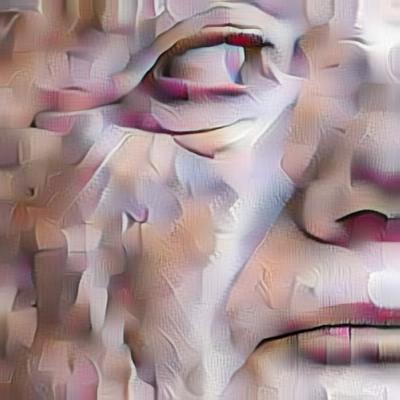}
    \includegraphics[width=0.24\linewidth]{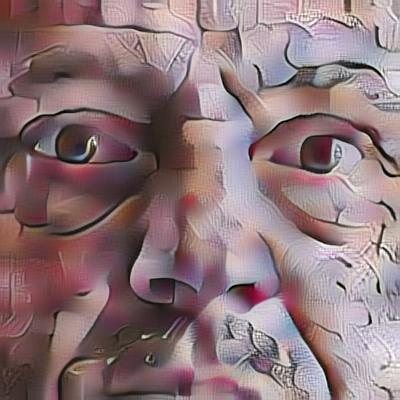}
    \includegraphics[width=0.24\linewidth]{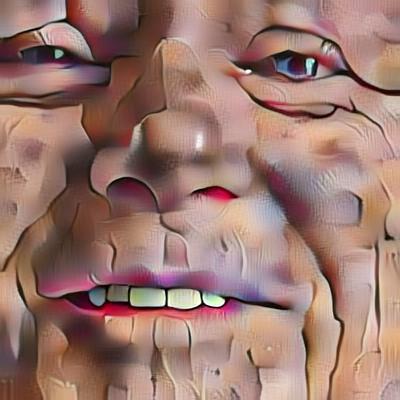}
    \includegraphics[width=0.24\linewidth]{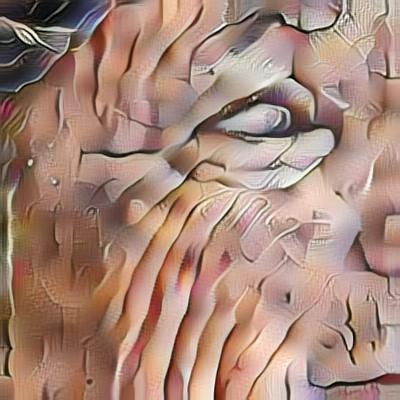}
    \includegraphics[width=0.24\linewidth]{images/instability_consistency_ini/homme_2.jpg}
    \includegraphics[width=0.24\linewidth]{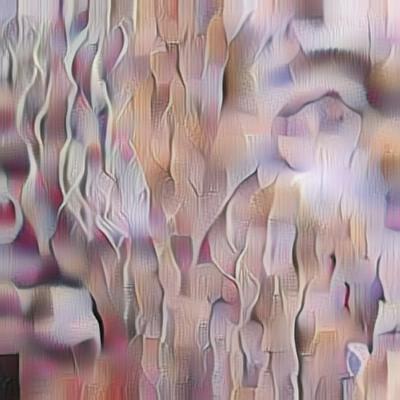}
    \caption{Showing the fourth iterates of sequence $(x_t)$ as defined in \eqref{eq:recursive_style_transfer} for MST method on various fragment of portraits. The tessellation effect is maintained across all the images.}
    \label{fig:consistent_first_iteration_divergence}
\end{figure}

A style transfer method is simply a function $f$ that takes as input a style image $s$ and a content image $c$ and outputs a stylized version $f(s,c)$ of $c$ with $s$.
It is reasonable when giving such a method the same image as content and style, to expect the image itself, \textit{i.e.} that $f$ satisfies $f(s,s)\approx s$.
Let us now consider the following recursion
\BEQ\label{eq:recursive_style_transfer}
x_{t+1} = f(x_{t}, x_{t})~,
\EEQ
where $x_0$ is an initial image.
Optimization based methods empirically converge to an equilibrium where $f(x, x) = x$ independently of the initialization.
On the opposite, feed-forward approaches to style transfer \citep{li2016precomputed,ulyanov2016texture,johnson2016perceptual,ghiasi2017exploring,li2017universal,huang2017arbitrary} lead to oscillating sequences $(x_t)$ around non-trivial (\textit{i.e.} not a monochrome image) forms, yet typically bearing absolutely no resemblance with the initial image $x_0$. 
Since the pixel values are clamped between 0 and 1, colours end up being either saturated or zero, but not uniformly and still revealing specific patterns in Figure \ref{fig:consistent_first_iteration_divergence} for instance. 
Interestingly also, when starting from very simple images $x_0$, like a uniform color, for some $f$, the sequence would still show the same type of instability in the long-run, see this video\footnote{https://youtu.be/WCJNLWb-H2M} for instance.

\begin{figure}[h!]
\centering
\includegraphics[width=0.24\linewidth]{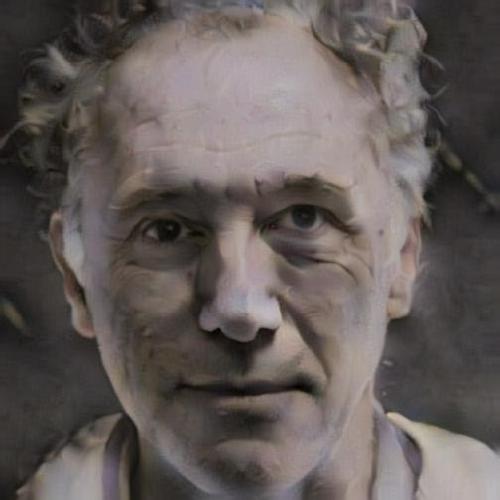}
\includegraphics[width=0.24\linewidth]{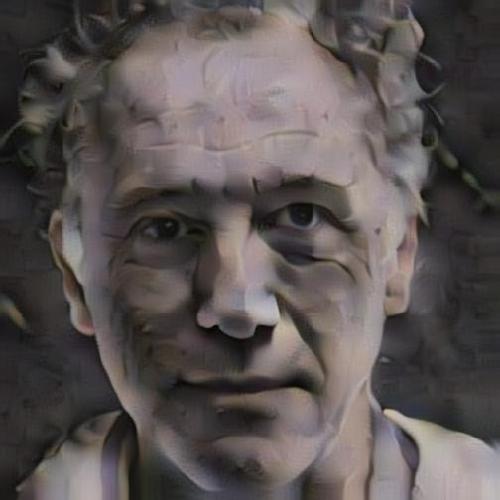}
\includegraphics[width=0.24\linewidth]{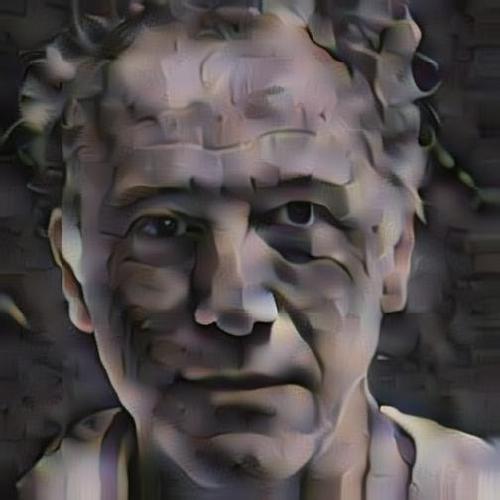}
\includegraphics[width=0.24\linewidth]{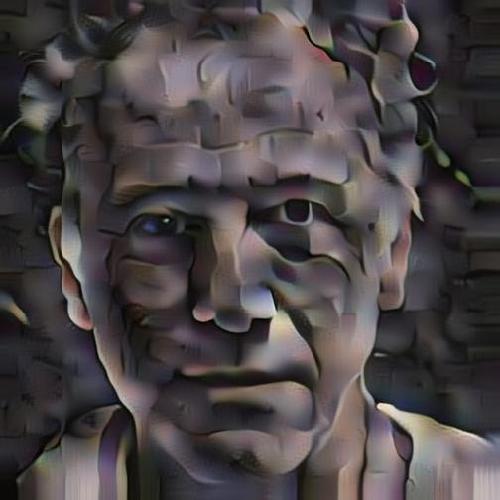}
\vspace{1mm}

\centering
\includegraphics[width=0.24\linewidth]{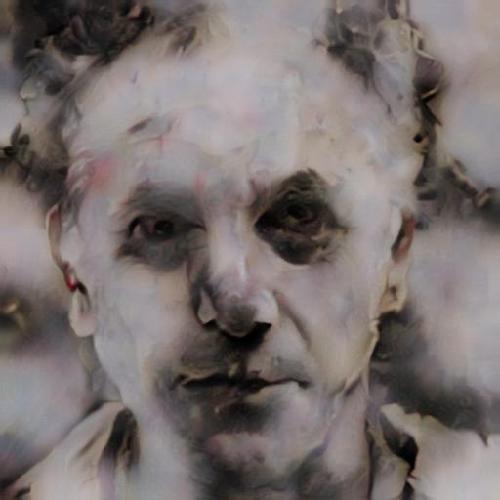}
\includegraphics[width=0.24\linewidth]{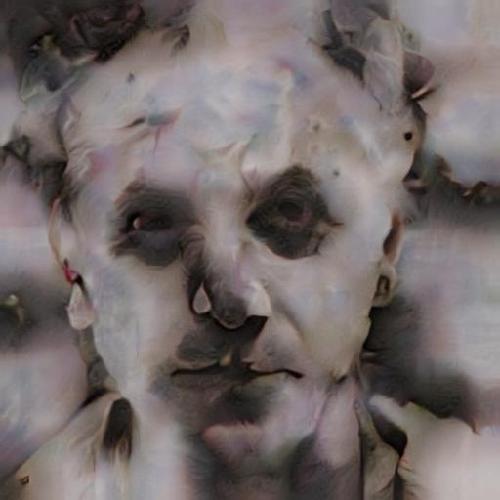}
\includegraphics[width=0.24\linewidth]{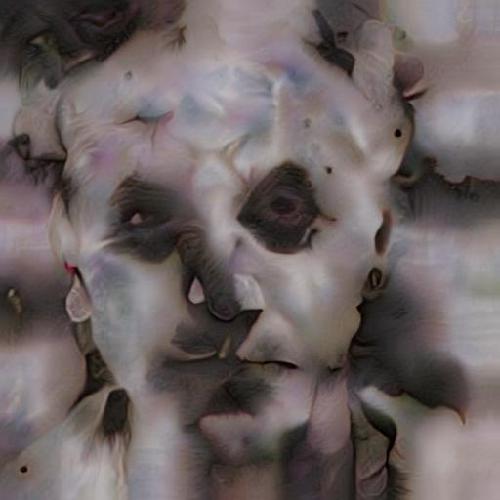}
\includegraphics[width=0.24\linewidth]{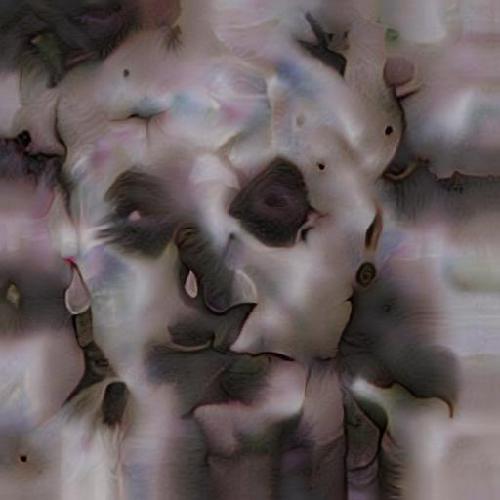}
\vspace{1mm}

\centering 
\includegraphics[width=0.24\linewidth]{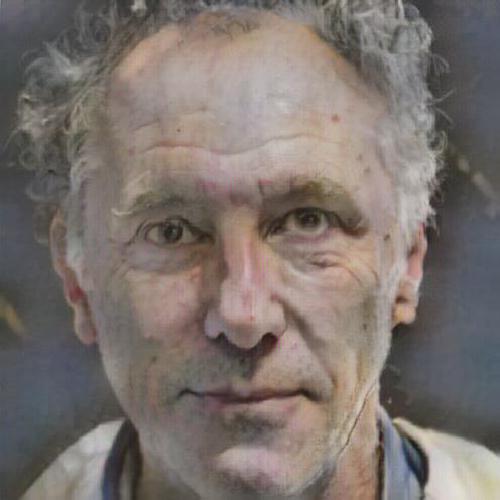}
\includegraphics[width=0.24\linewidth]{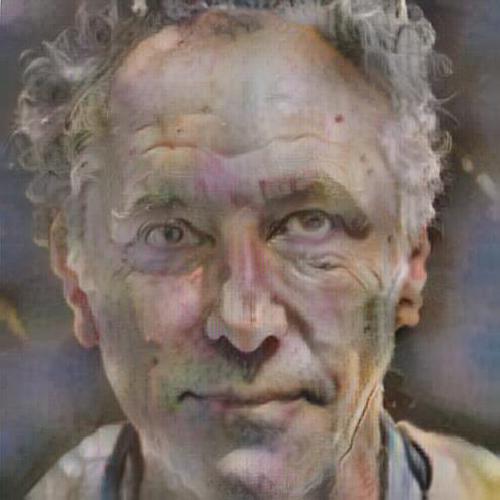}
\includegraphics[width=0.24\linewidth]{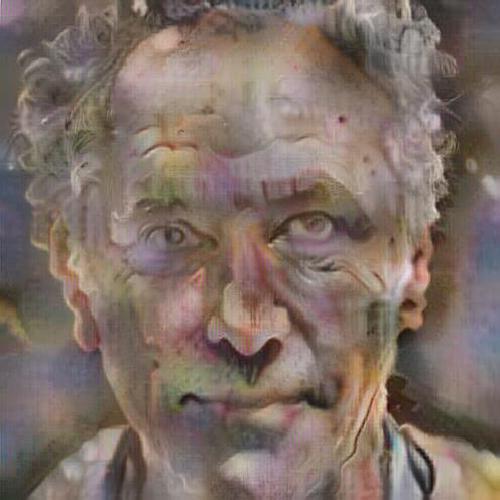}
\includegraphics[width=0.24\linewidth]{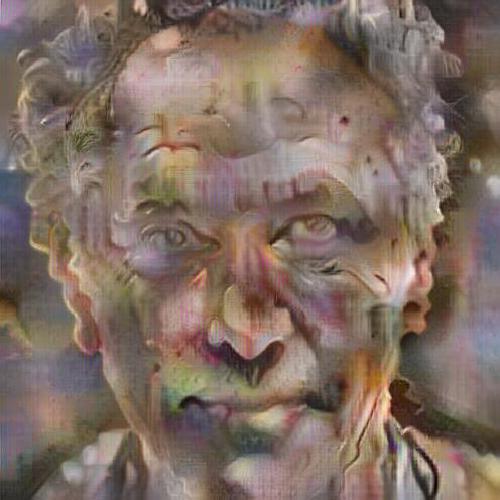}

\caption{Each row represents the first fourth iterate of sequence defined in \eqref{eq:recursive_style_transfer} with a different neural style transfer approach. The first row corresponds to MST \citep{zhang2019multimodal}, the second to WCT \citep{li2017universal} and the third  to AdaIn \citep{huang2017arbitrary}.
The first method has a clear tessellation effect, the second has a blurring effect and the third gives an increasingly clownish style to the image. These effects appear to be the same across images, see Figure \ref{fig:consistent_first_iteration_divergence}.}
\label{fig:repetition_instability}
\end{figure}

From the perspective of computational creativity, this seeming failure is interesting.
In the first iterations, we observe that some methods produce a series of images progressively stylized.
Given a style transfer function $f$, the very same effect happens across all sequences we experimented with.
For instance, in Figure \ref{fig:repetition_instability} we see a distinct tessellation effect in the images on the first row.
We use this technique in the interactive painting experiments to produce more diverse and computationally creative style transfer outputs, see image (f) in Figure \ref{fig:iterative_same_portrait} for instance.

\begin{figure}[h!]
\centering
\includegraphics[width=0.24\linewidth]{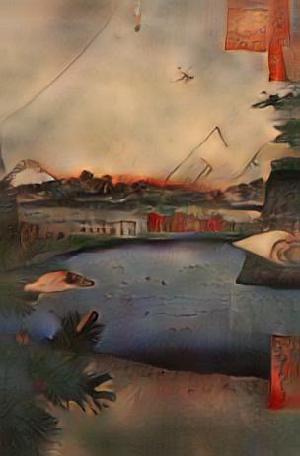}
\includegraphics[width=0.24\linewidth]{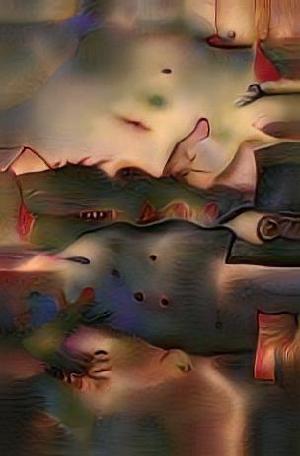}
\includegraphics[width=0.24\linewidth]{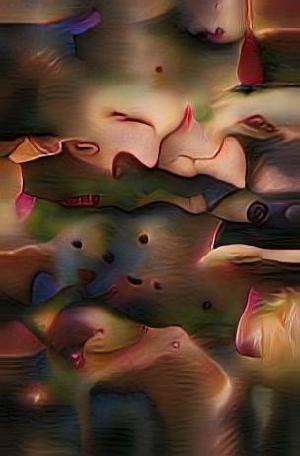}
\includegraphics[width=0.24\linewidth]{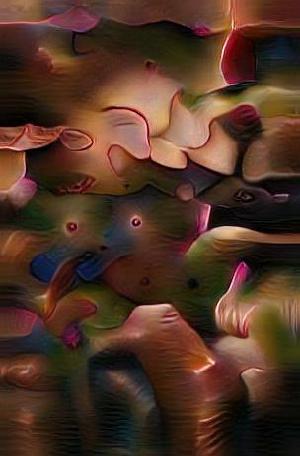}

\includegraphics[width=0.24\linewidth]{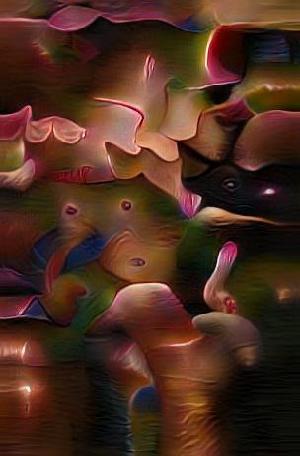}
\includegraphics[width=0.24\linewidth]{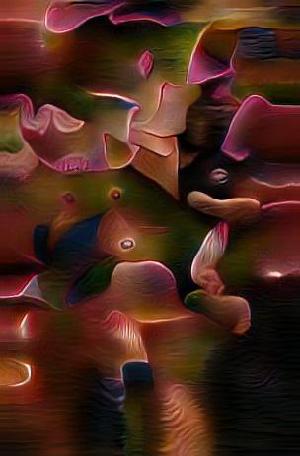}
\includegraphics[width=0.24\linewidth]{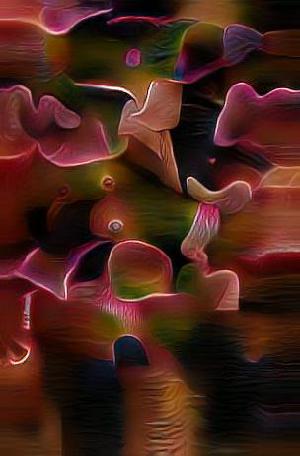}
\includegraphics[width=0.24\linewidth]{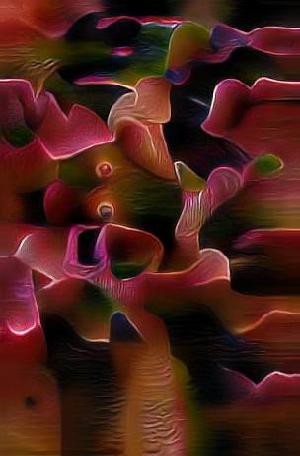}

\includegraphics[width=0.24\linewidth]{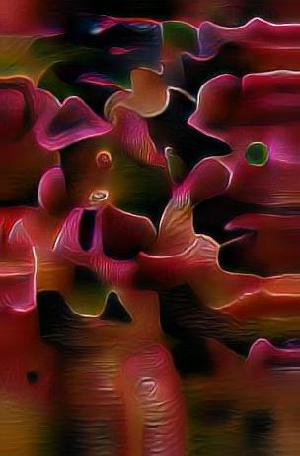}
\includegraphics[width=0.24\linewidth]{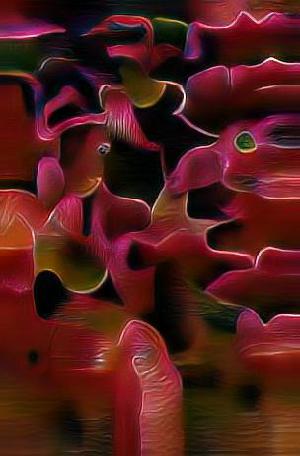}
\includegraphics[width=0.24\linewidth]{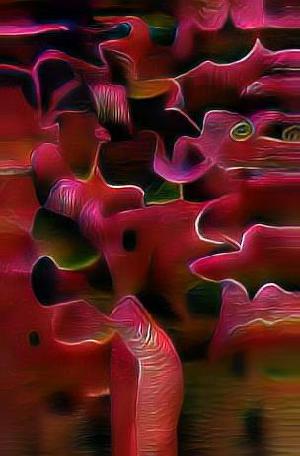}
\includegraphics[width=0.24\linewidth]{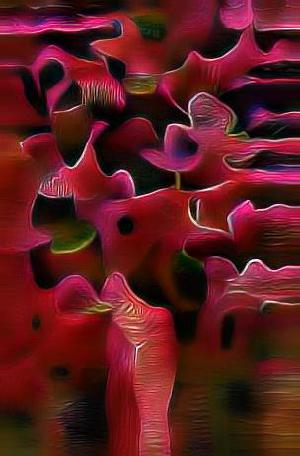}

    \caption{We observe the asymptotic regime of \eqref{eq:recursive_style_transfer} as the images represents the first $(x_{4k})_{k=0,\ldots,11}$ iterations of \eqref{eq:recursive_style_transfer} for a given initial image.}
    \label{fig:frame_divergence}
\end{figure}

Alternatively the asymptotical regime of the sequences $(x_t)$ produces surprising animations.
The appearing patterns are completely different from one approach to another, but are experimentally the same for different initialisation images and a given method.
Sequences are shown in Figure \ref{fig:frame_divergence}, refer to this \href{https://youtu.be/gAq1lvb1G1c}{video}\footnote{https://youtu.be/gAq1lvb1G1c}  or this \href{https://youtu.be/s87R-9JITvE}{one}\footnote{https://youtu.be/s87R-9JITvE} for a more lively visualisation.

\section{Interactive Painting Experiments}\label{sec:interactive_paintings}

In the previous study, we have questioned the relevance of neural style transfer evaluation.
To go beyond comparing techniques, we propose to take advantage of the diversity of the outputs and to use them as a source inspiration for artists. 

Some painters have recently explored interactive processes with machines in the real world, particularly in the case of painting.
For instance, \cite{chung2015drawing} among others, leveraged on the algorithms from artificial intelligence to paint interactively with humans in the real world, where a machine would act on the real canvas via a robotic arm.
\cite{DialogCanvasMachine} also explore such an interaction, where the machine does not act but suggests via projection. 
However, none of these use style transfer algorithms outputs to paint interactively with an artist on the canvas.

We explore that possibility through various series of interactively painted portraits.
We describe here various interactive painting experiments inserting outputs of neural style transfer algorithms during the human painting process.
In all cases, the algorithms' creations are projected onto the canvas but never automatically painted, for instance via a robotic arm or a printer.
We first describe the experiments on canvas and then motivate the underlying design choices.
We finally show how the notion of \textit{computational catalyst} naturally emerges. Note also that all the paintings revolve around portrait themes.

\subsection{The portraits}

\begin{figure}
    \centering
    
    \begin{minipage}{.475\linewidth}
    \includegraphics[width=\linewidth]{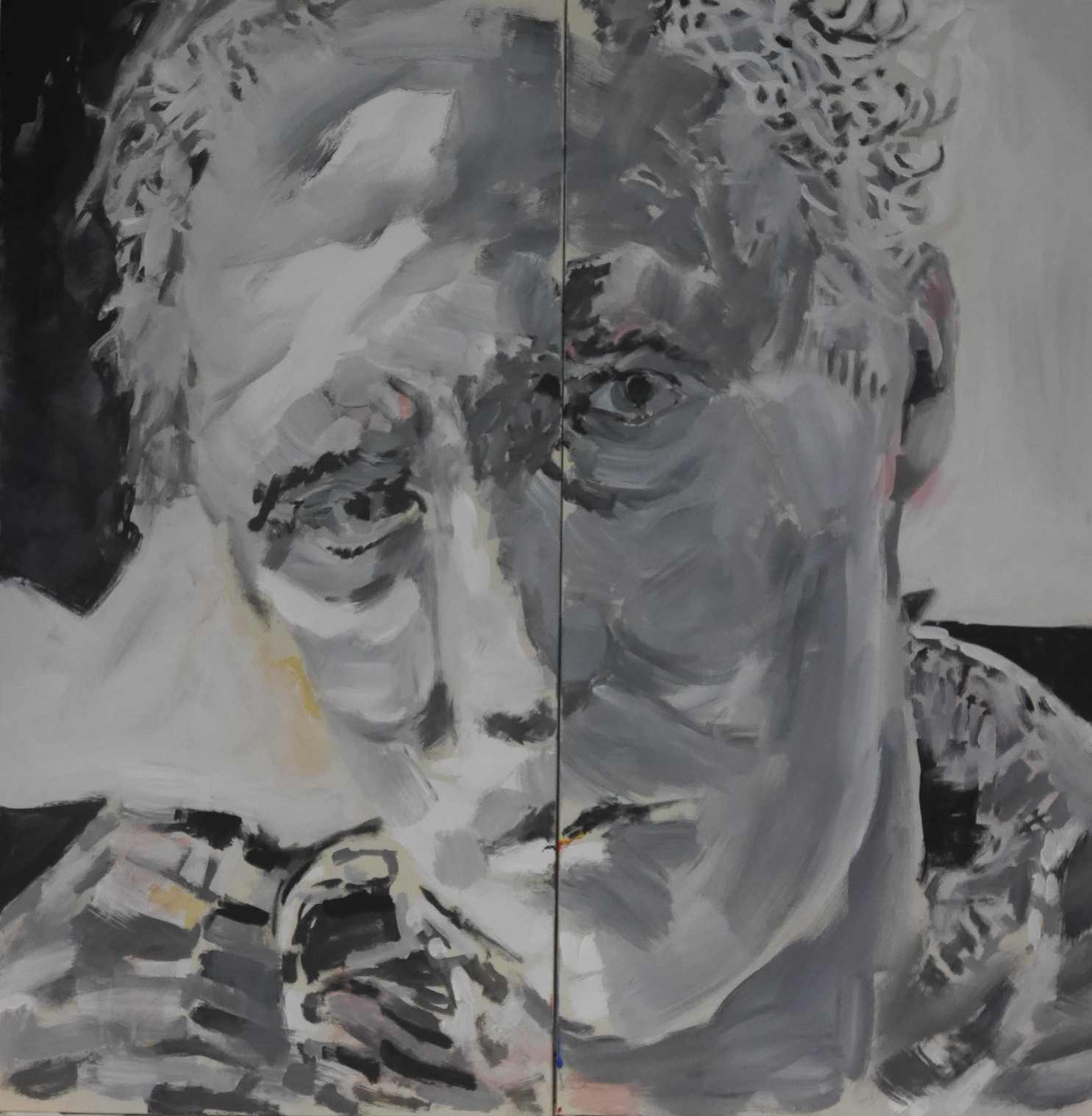}
    \end{minipage}
    \begin{minipage}{.49\linewidth}
    \includegraphics[width=0.49\linewidth]{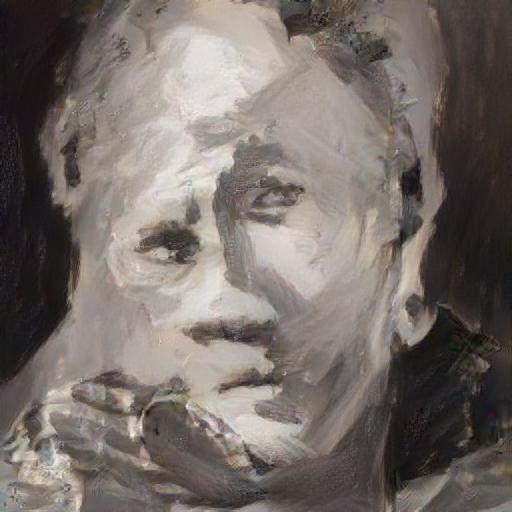}
    \includegraphics[width=0.49\linewidth]{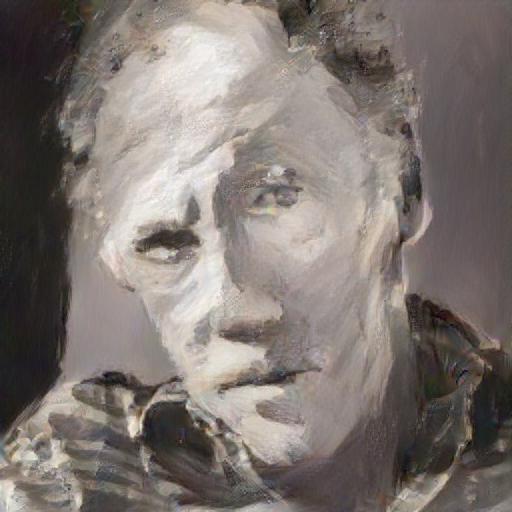}\\
    \includegraphics[width=0.49\linewidth]{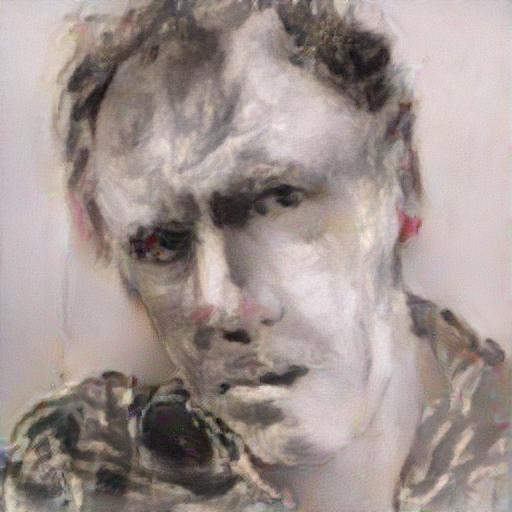}
    \includegraphics[width=0.49\linewidth]{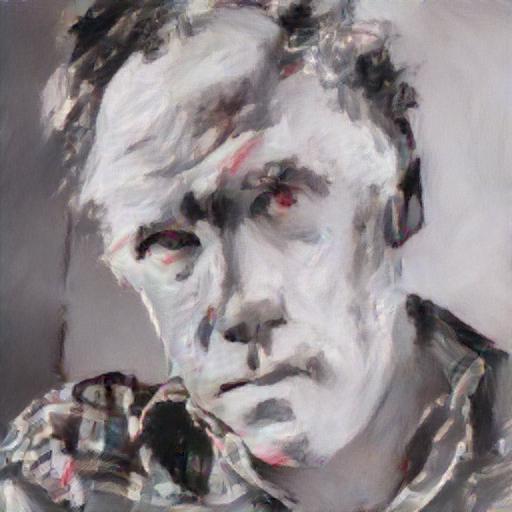}
    \end{minipage}
    
    \caption{Left: two panels of 100$\times$50 oil canvas. Right: the style transfer outputs that formed the painter theme and were successively projected on the canvas. For each, the three canvas in Figure \ref{fig:thomas_portrait_multiple} were used as style images; the content was a photographic portrait.}
    \label{fig:many_outputs}
\end{figure}

\paragraph{Editing multiple styles in one portrait.}
Neural style transfer outputs are very diverse from one method to another, as outlined in Figure \ref{fig:style_transfer_ex} for instance.
To edit the creative content of these outputs into a single final artwork, we select a person's photography as a content image and we transfer the style of previous artists' painting into this content image using various algorithms.
We then show the stylized images to the artist, but not the original content image, which he never sees.
He then selects some of the outputs that best resonate with his practice.
These style transfer outputs are finally alternatively projected on a canvas for a certain amount of time.
Figure \ref{fig:many_outputs} is an example of canvas realized according to this process; we complement the final canvas with the various style transfer outputs chosen by the artist. We also explore other variations of that idea, for instance via collage, where the selected outputs are mingled together into a single image which is projected afterwards.

\begin{figure}
\begin{minipage}{.51\linewidth}
\includegraphics[width=\linewidth]{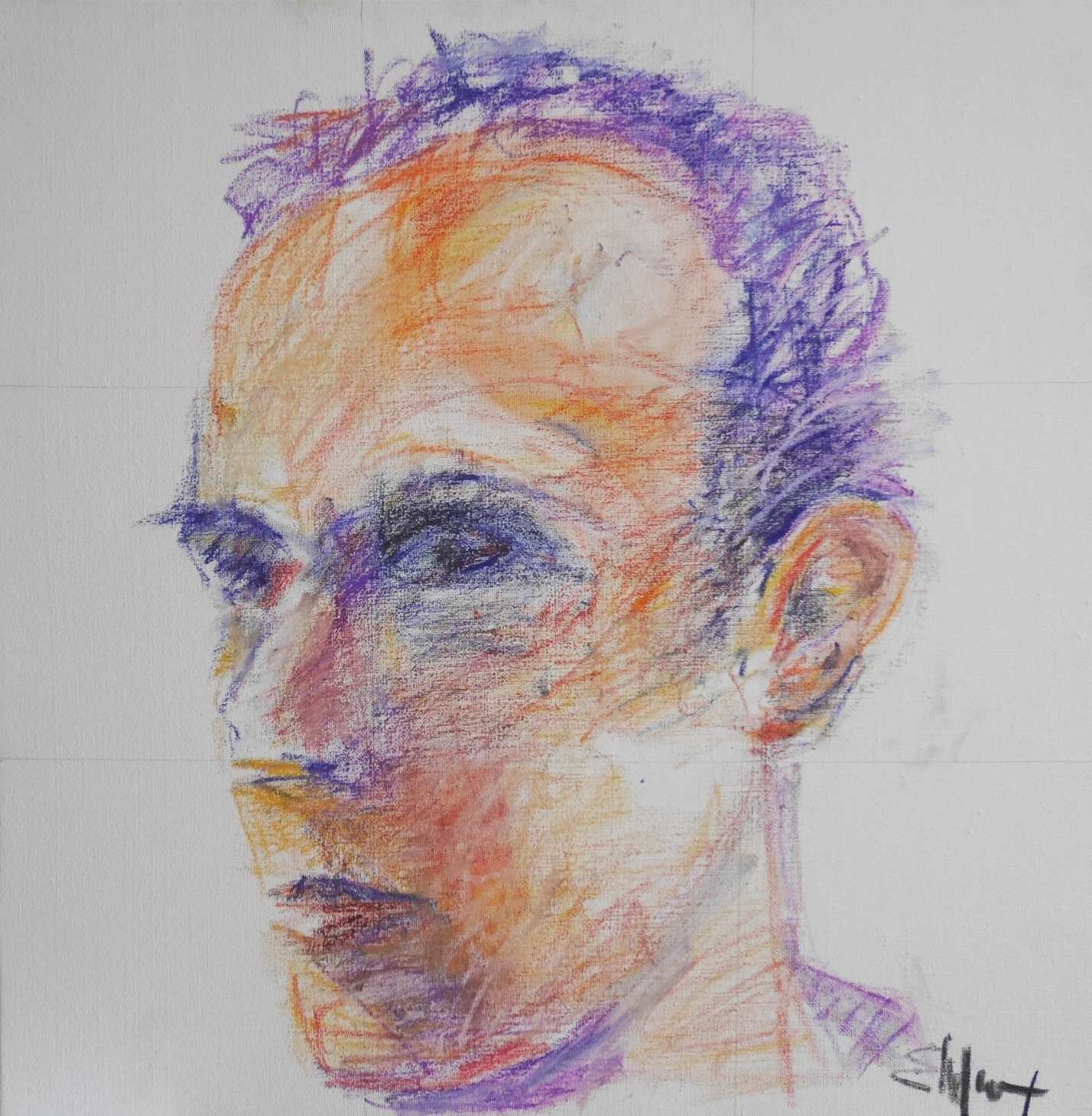}
\end{minipage}
\begin{minipage}{.48\linewidth}
\includegraphics[width=0.49\linewidth]{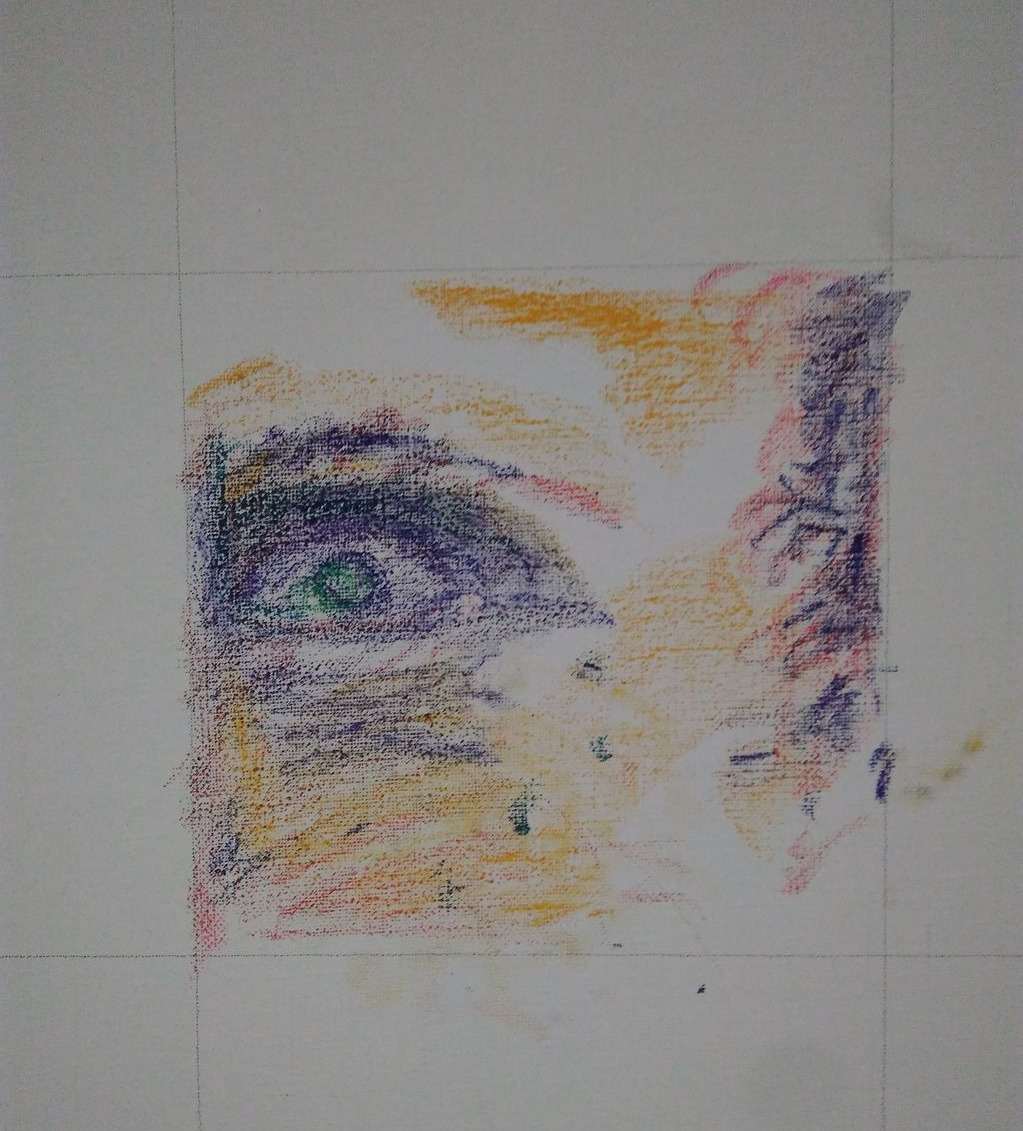}
\includegraphics[width=0.49\linewidth]{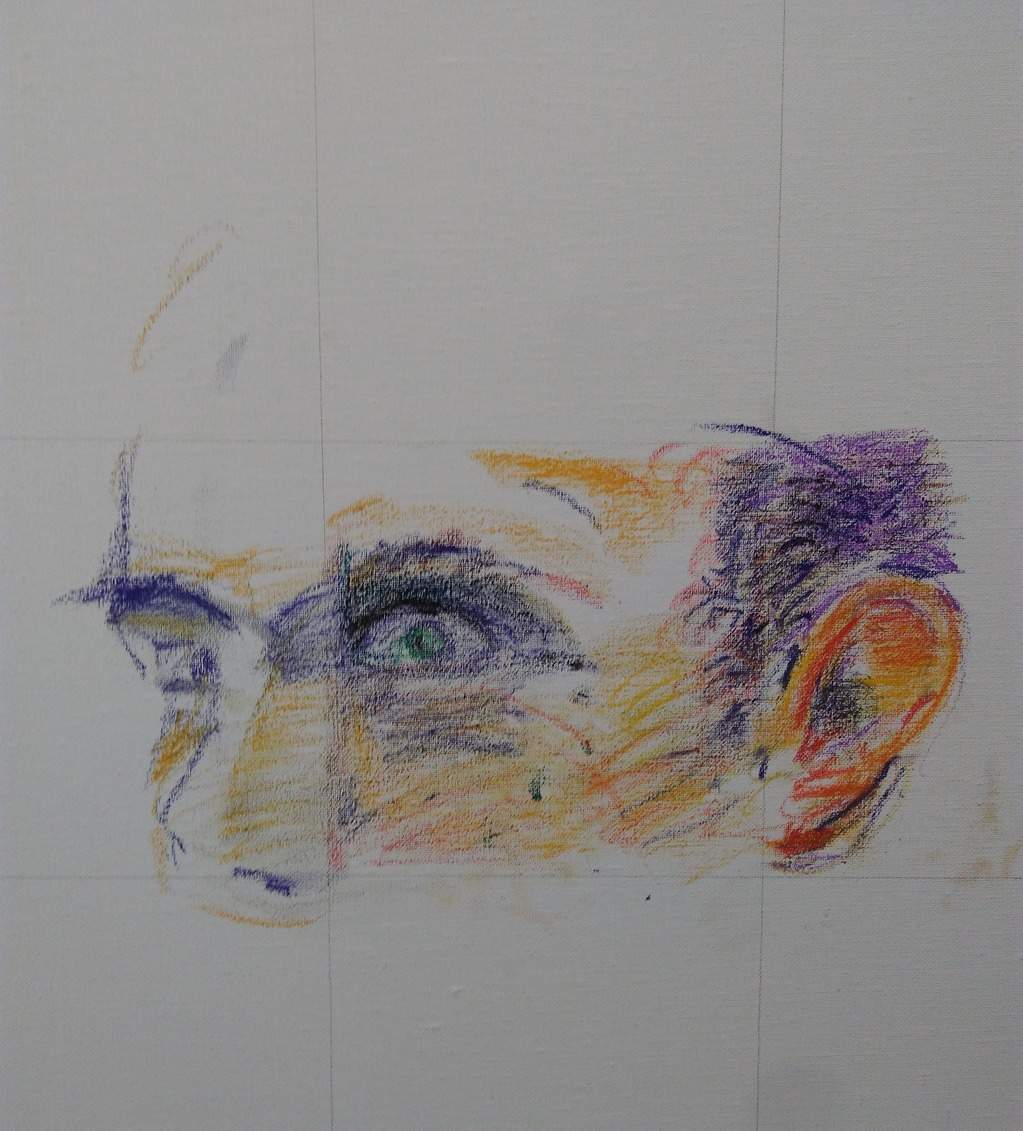}\\
\includegraphics[width=0.49\linewidth]{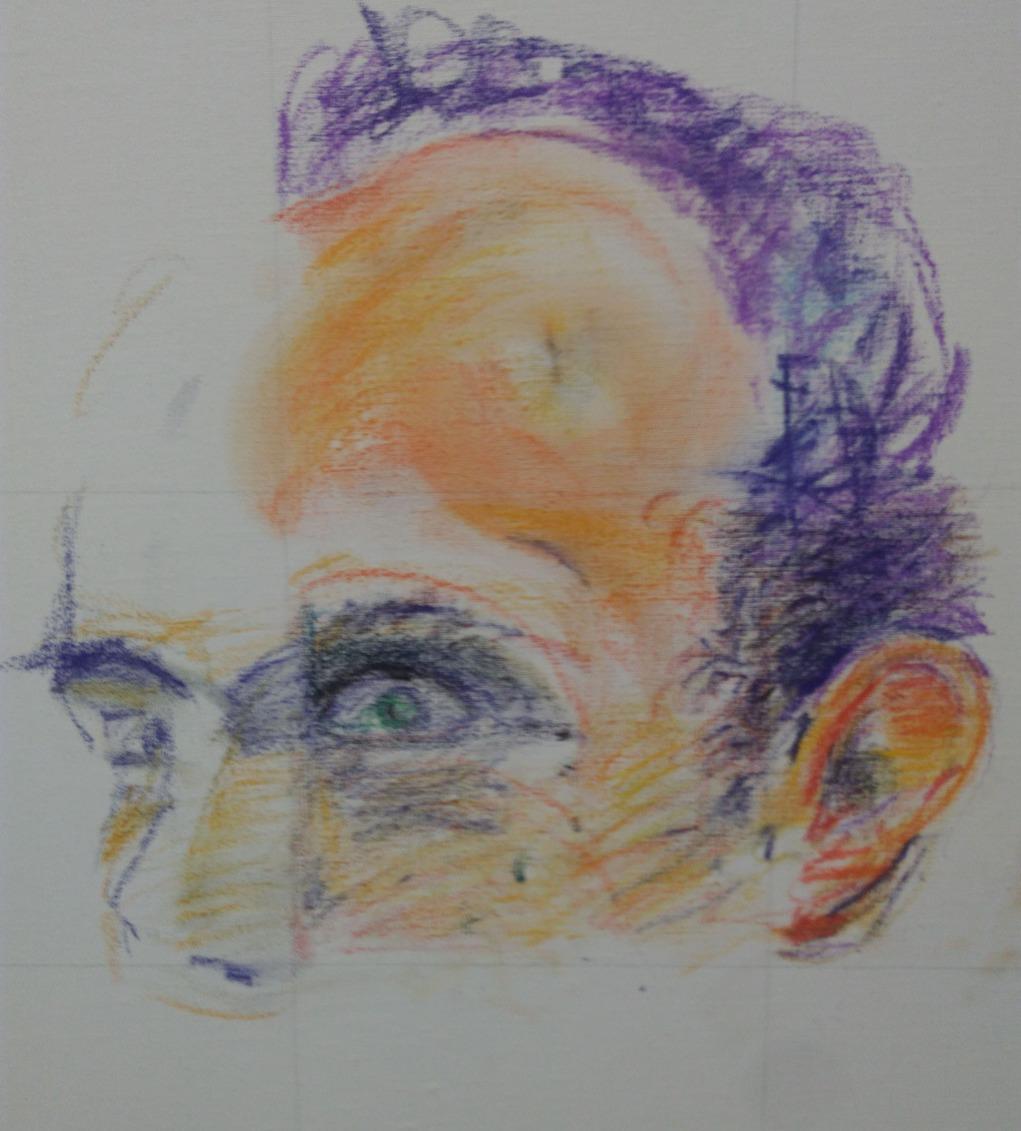}
\includegraphics[width=0.49\linewidth]{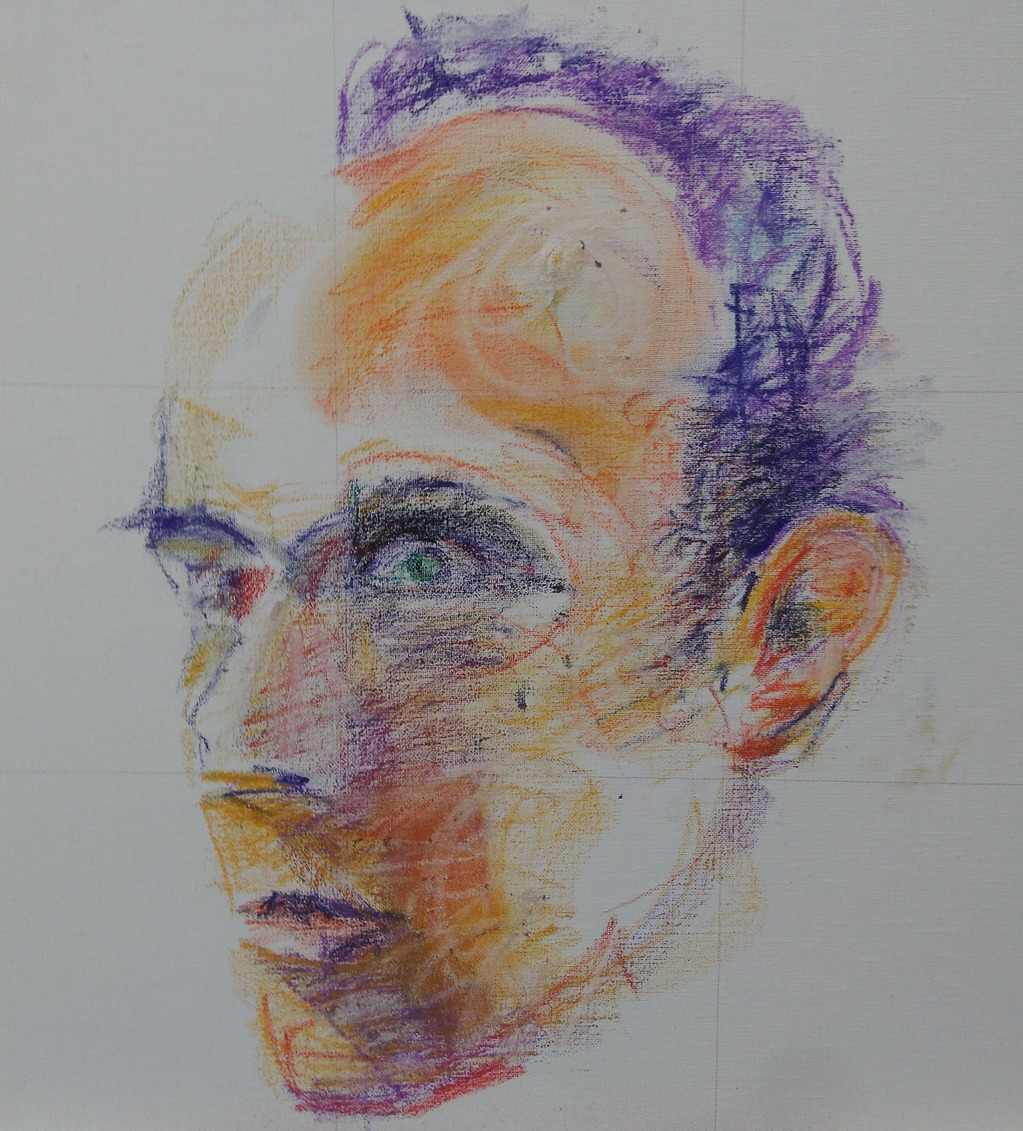}
\end{minipage}
\caption*{(left) final canvas (right) steps 1, 3, 5, 7}

\hspace{8mm}

\subfloat {{\includegraphics[width=0.195\linewidth]{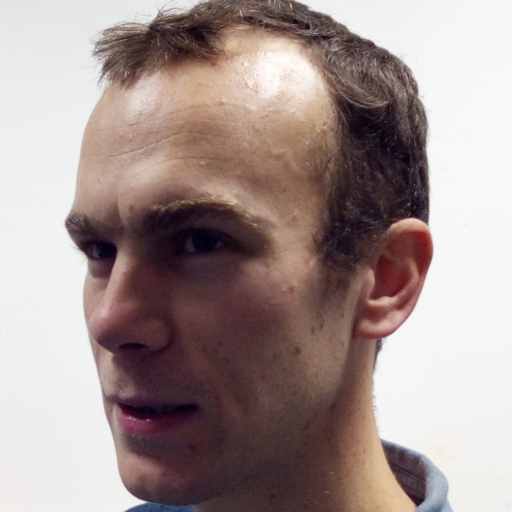} }}
\subfloat {{\includegraphics[width=0.195\linewidth]{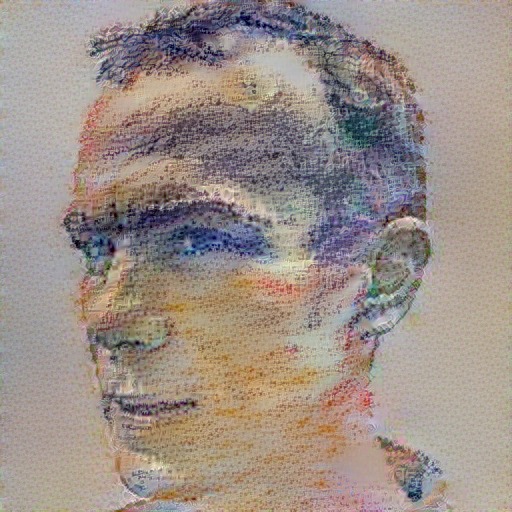} }}
\subfloat {{\includegraphics[width=0.195\linewidth]{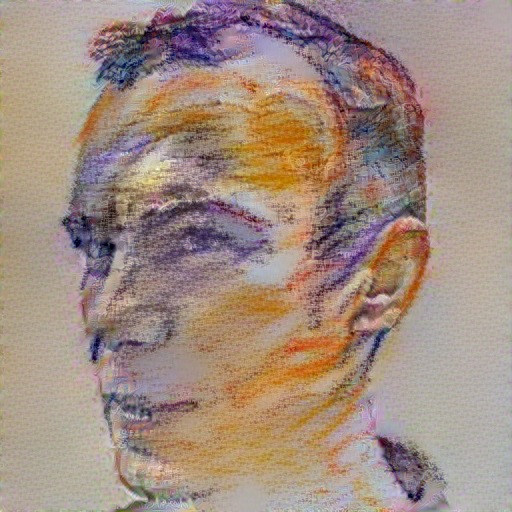} }}
\subfloat {{\includegraphics[width=0.195\linewidth]{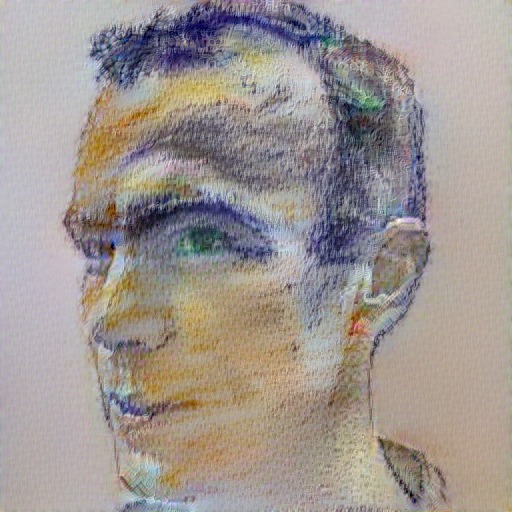} }}
\subfloat {{\includegraphics[width=0.195\linewidth]{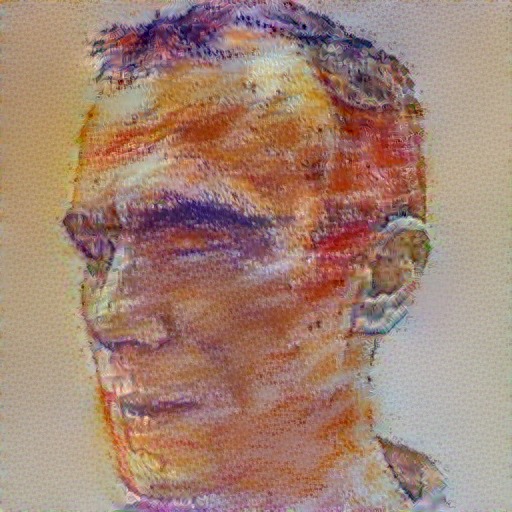} }}
\caption*{Original photographic and projections after steps 1, 3, 5, 7.}

\caption{ A 50$\times$50 oil pastel canvas inspired by an evolving projection. 
The canvas was divided into 9 squares and the painter had paint sequentially on each square.
After each square was completed, the output of the style transfer algorithm using the original photographic as a content and the current canvas as a style image was projected.
}
\label{fig:segmented_portrait}
\end{figure}

\paragraph{Pixelizing portrait construction.} 
The motivation of this creative process is to artificially create an interactive loop between the painter and the algorithm.
The initial image is projected onto or next to the canvas, which is divided into squares.
The painter is then asked to paint sequentially on each square of the canvas.
Whenever a square is completed, we use it as a new style image to stylize the initial photographic-like image of the portrait, which is then projected on the canvas, images (a)-(d) in Figure \ref{fig:segmented_portrait} are some of these projected outputs. Anytime then, the painter only sees an interpretation of the photographic portrait by a style transfer algorithm taking the painter's style in the previous painting. We show one example of a canvas produced in such a way in Figure \ref{fig:segmented_portrait}.

Note that the style transfer output should be the machine's prediction of what the artist would do, provided at least that the previous square contains all the style information of the painter and that the style transfer method is ideal. This remark was then the basis for a gamification exploration of the painting process, where the artist was asked to attempt to fail the machine prediction as much as possible.

This decomposition of the painting process produces painting artworks that are sequential objects. Not only the final canvas is interesting but all its part.
Actually, algorithms in image computational creativity are very much less performative at generating images as sequences of brushstrokes as they are at generating images all at once, like with GANs \citep{goodfellow2014generative} for instance. This is simply because paintings are usually not sequential objects.
Indeed we very rarely observe all the steps leading to a painting, apart for large quantities and categories of simple sketches \citep{eitz2012hdhso,quickDraw}.
Alternatively, computationally inferring the steps of a painting from the final canvas \citep{xie2013artist,ganin2018synthesizing,nakano2019neural} is not yet very successful.
This arguably explained why in painting, compared to other domains such as music, whose artworks are sequential by nature, the computationally creative algorithms are harder to frame in a fully interactive way with humans, hence limiting the ability for a painter to truly interact with machines.

\paragraph{Interactive series of portraits.} 

\begin{figure}
\subfloat[First canvas]{{\includegraphics[width=0.33\linewidth]{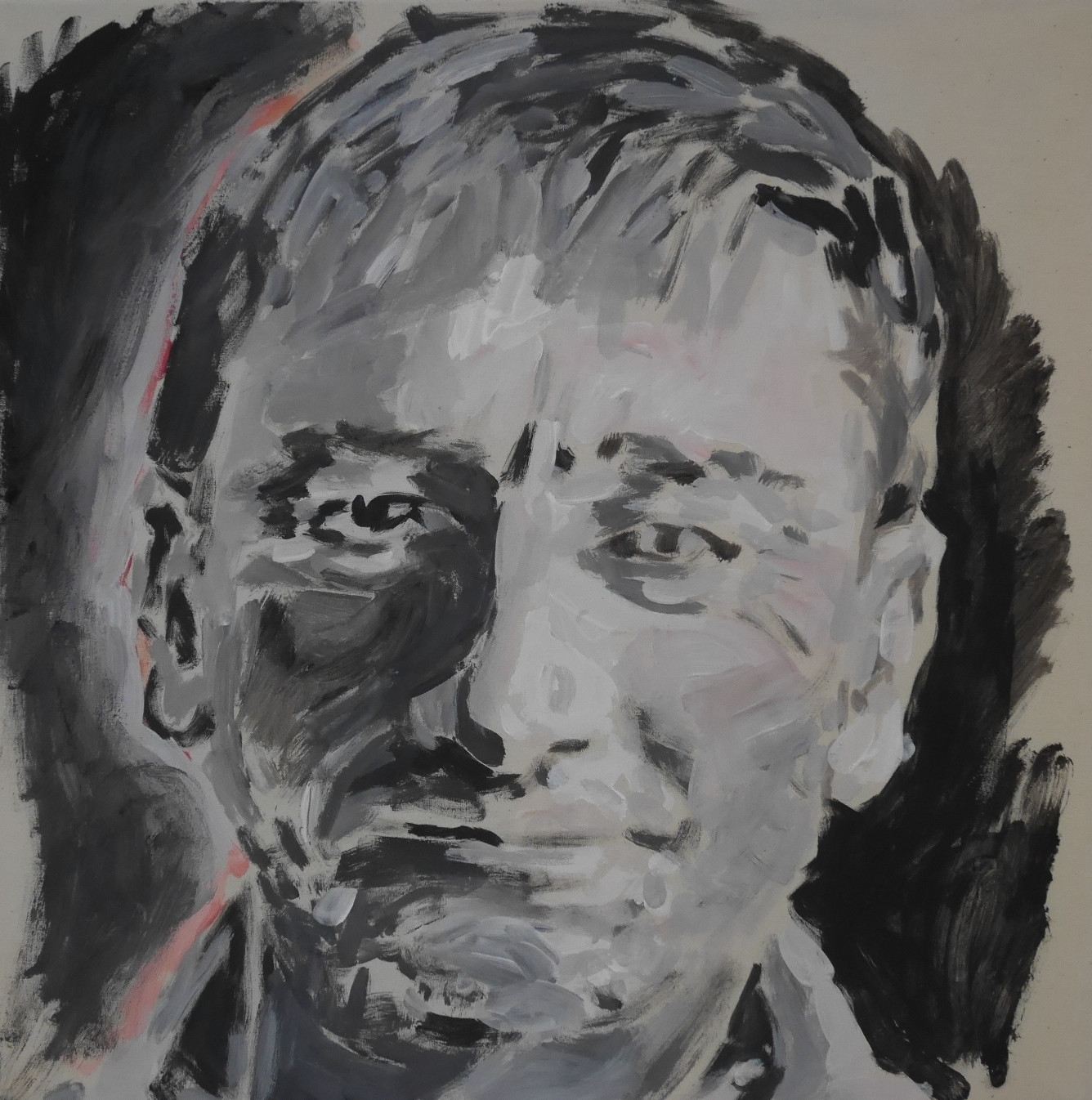} }}
\subfloat[Second canvas]{{\includegraphics[width=0.33\linewidth]{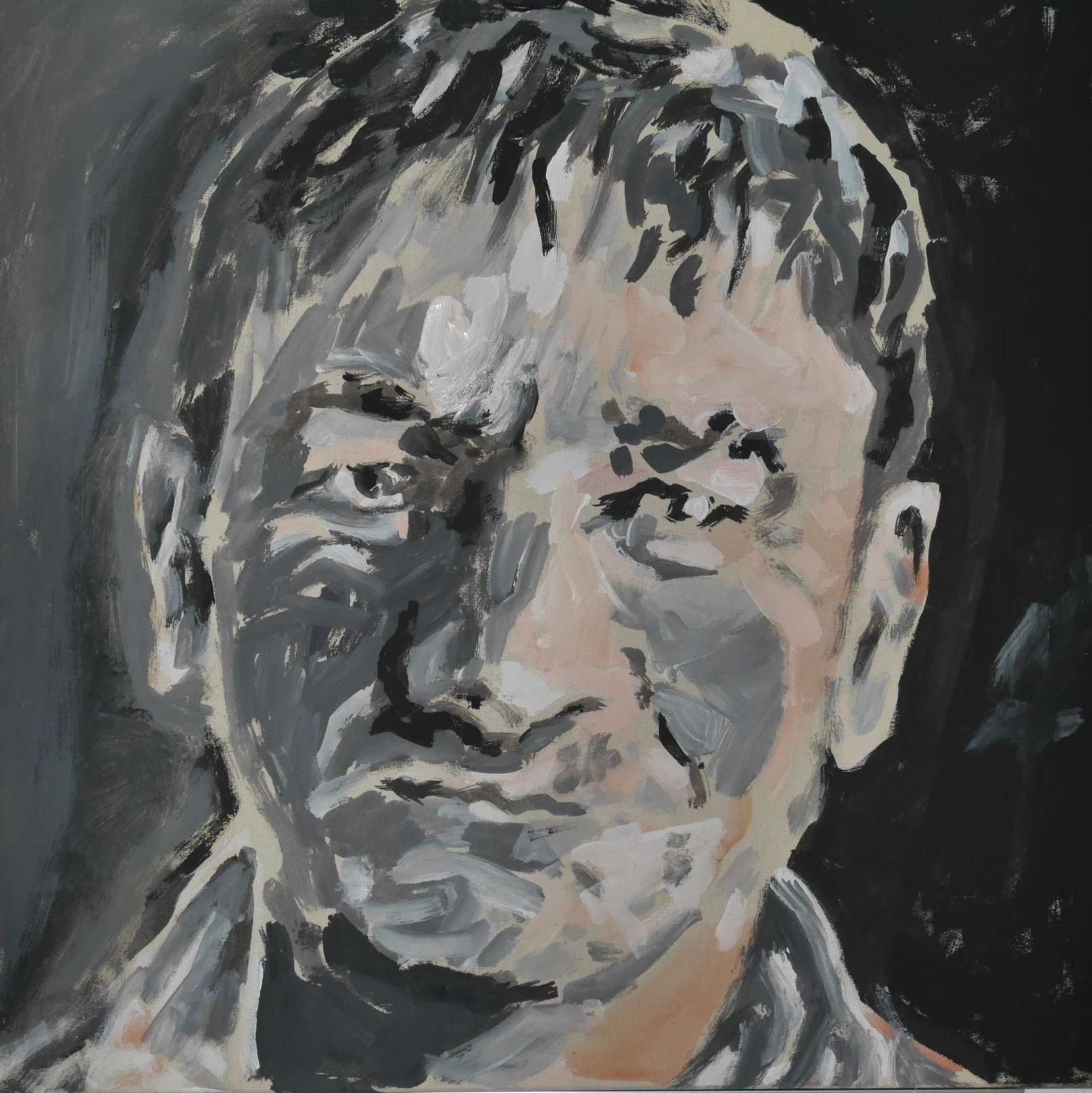} }}
\subfloat[Third canvas]{{\includegraphics[width=0.31\linewidth]{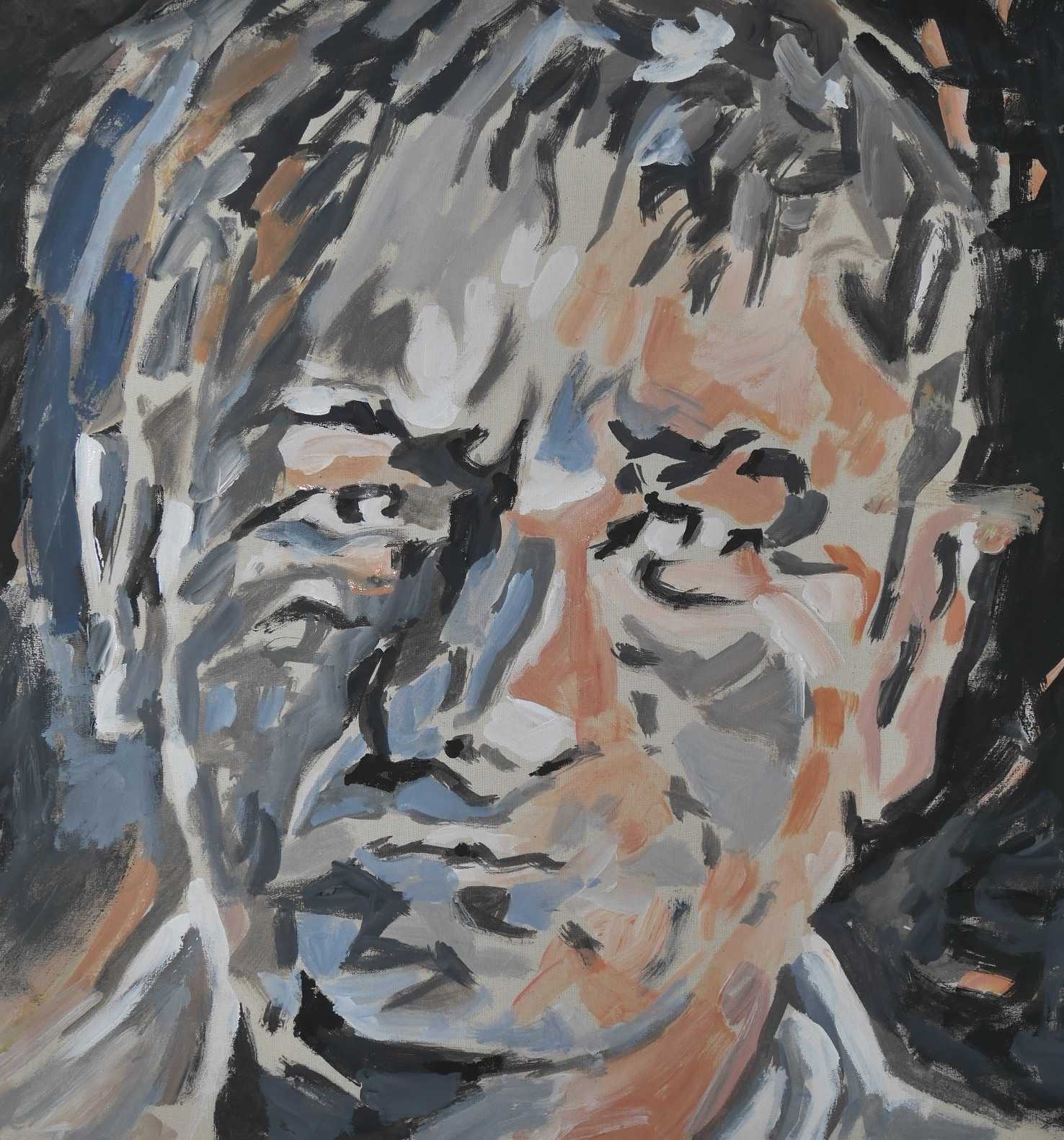} }}

\subfloat[Projection $1$]{{\includegraphics[width=0.3\linewidth]{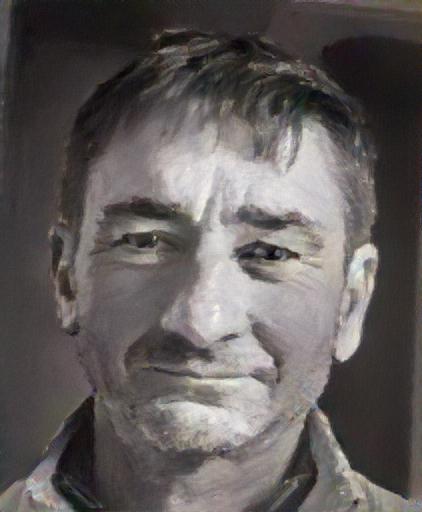} }}
\subfloat[Projection $2$]{{\includegraphics[width=0.3\linewidth]{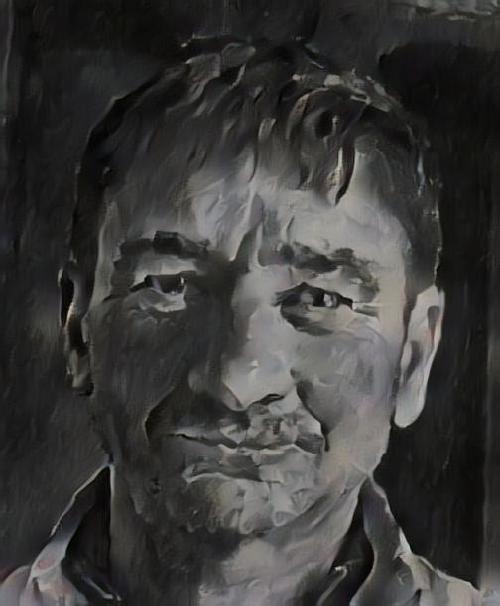} }}
\subfloat[Projection $3$]{{\includegraphics[width=0.365\linewidth]{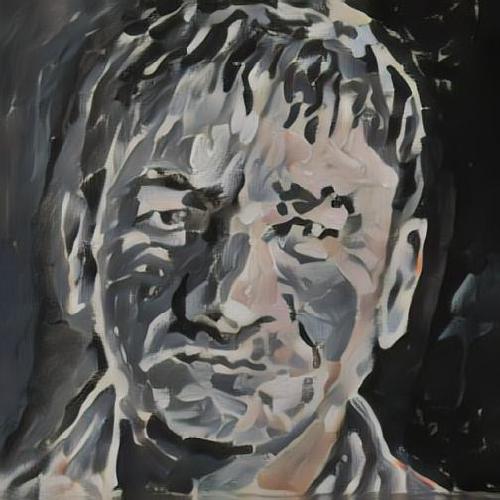} }}

\caption{Three 50$\times$50 cm oil on canvas of the same face. Iteratively showing style re-interpretation to the painter. Image (a) served as a style image to produce (e); Image (b) served as a style image to produce (f). Note that outputs (f) is the third iterate of \eqref{eq:recursive_style_transfer} with the MST style transfer algorithm, in order to produce a slight tessellation effect. }
\label{fig:iterative_same_portrait}
\end{figure}

\begin{figure}
\centering
\subfloat[First canvas.]{{\includegraphics[width=0.3\linewidth]{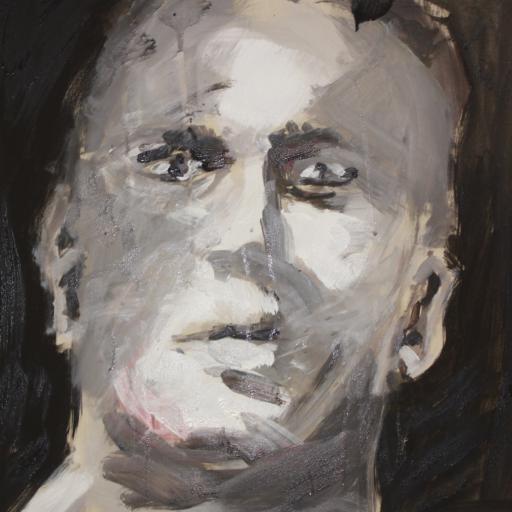} }}
\subfloat[Second canvas.]{{\includegraphics[width=0.3\linewidth]{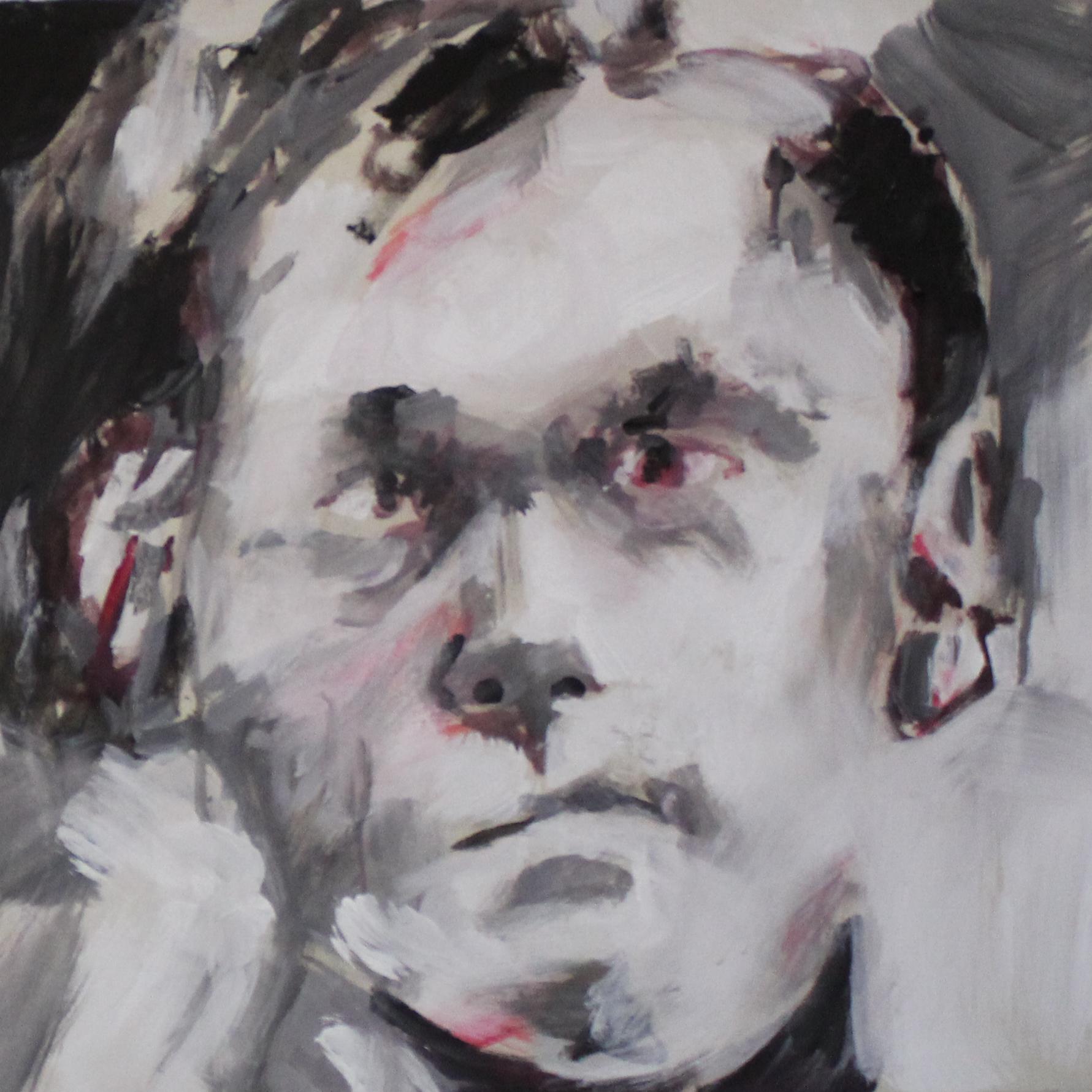} }}
\subfloat[Third canvas.]{{\includegraphics[width=0.3\linewidth]{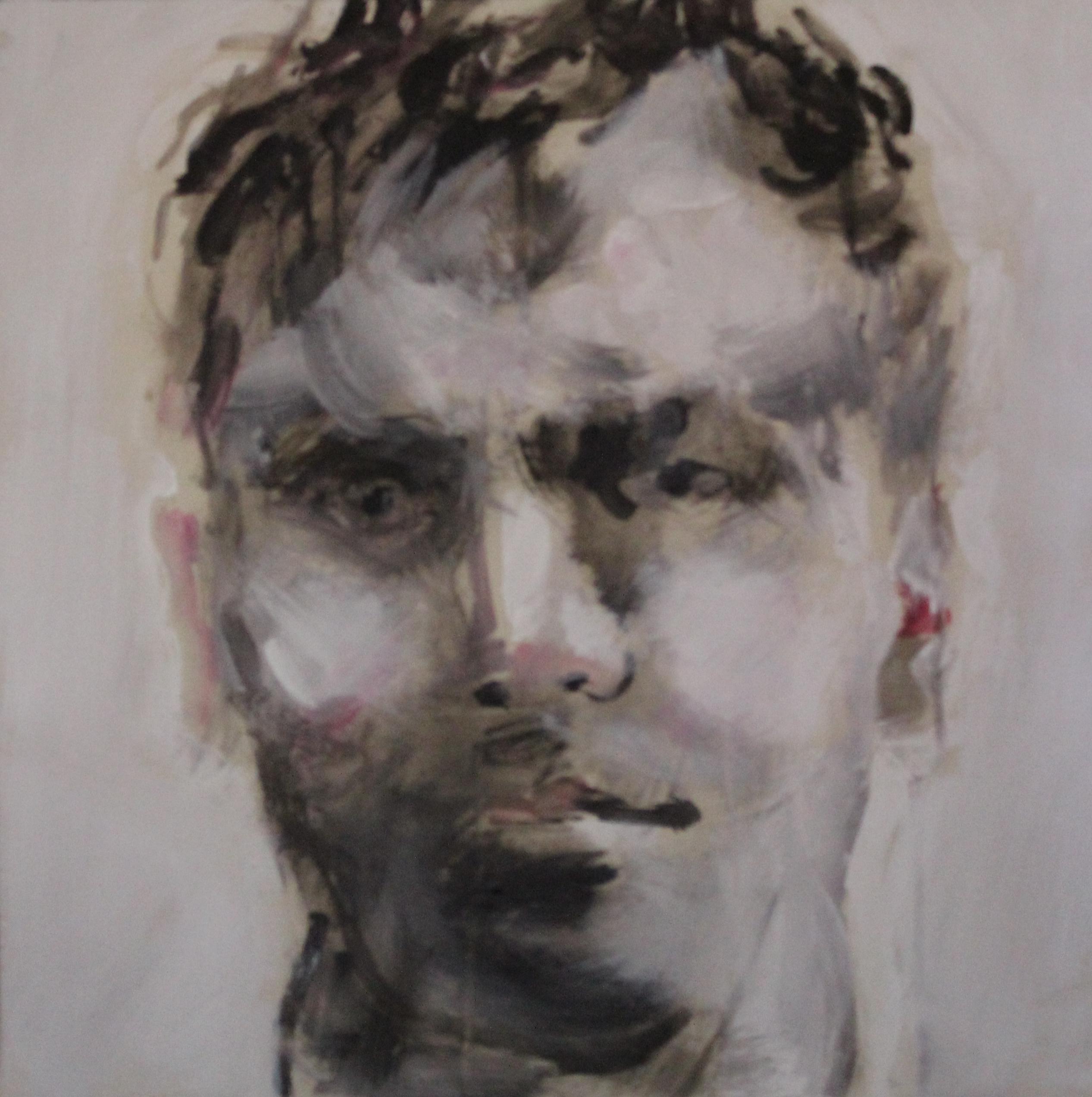} }}\\
\hfill

\centering
\subfloat[NST output 1]{{\includegraphics[width=0.3\linewidth]{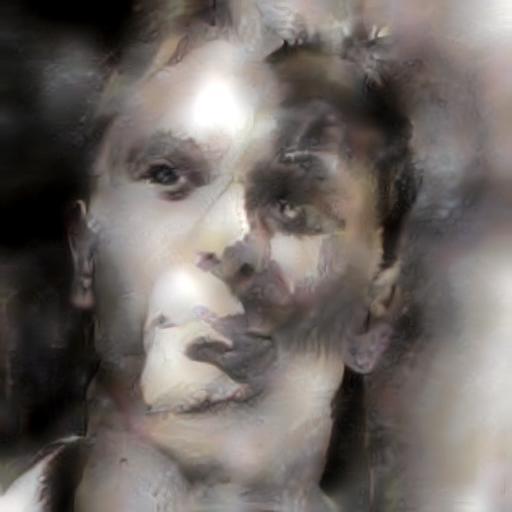}}}\vspace{1mm}
\subfloat[NST output 2]{{\includegraphics[width=0.3\linewidth]{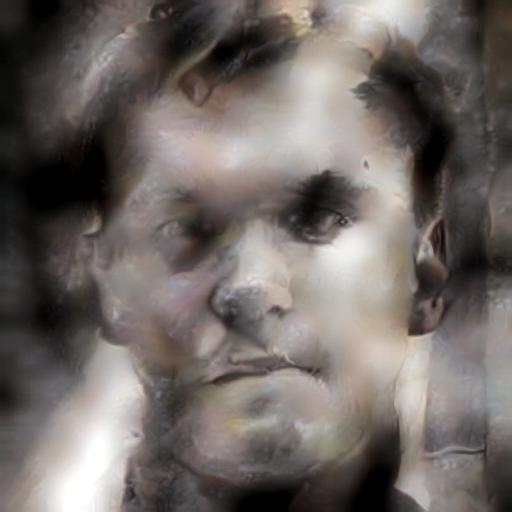}}}\vspace{1mm}
\subfloat[NST output 3]{{\includegraphics[width=0.3\linewidth]{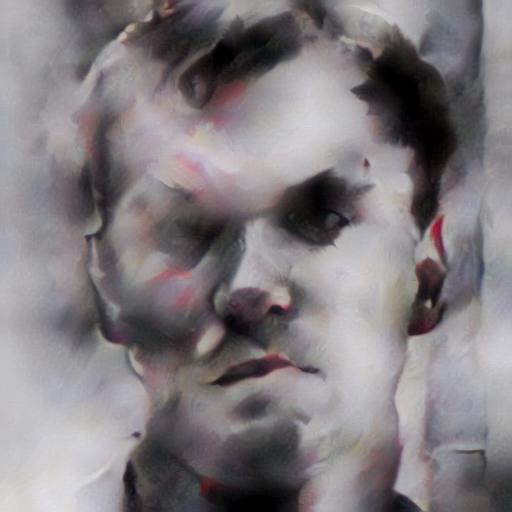}}}

\caption{The top row collects pictures of the paintings, 50$\times$50 cm oil on canvas. The bottom row gathers the outputs of the neural style transfer (NST) methods which were chosen by the painter as his basis theme. The first is the stylization of the photographic image with a previous painting of the painter. The second one is the stylization of a different photographic image with the first painting as a style image. The third is the stylization of the same previous photographic image with the second painting as a style image. Note that the order of the painting is chronological.}
\label{fig:thomas_portrait_multiple}
\end{figure}

We then considered using neural style transfer outputs for series of portraits.
We select a photographic portrait and we stylize with a neural style transfer algorithm it using a previous artists' painting as a style image.
We project the stylized image as inspiration for the painter.
When the painter has finished the painting, we stylize again the photographic portrait using the painting that has just been painted.
We project the new stylized image as the next inspiration, and we repeat the process, typically two or three times.
Figure \ref{fig:iterative_same_portrait}-\ref{fig:thomas_portrait_multiple} present two series of canvas in chronological order.
In figure \ref{fig:iterative_same_portrait} all canvas stem from the same photographic portrait, while in Figure \ref{fig:thomas_portrait_multiple} we alternate between two photographic portraits to avoid specialization of the artist to particular content.

We played on the many ways to generate new style images at each iteration. Importantly, at each iteration, the previous image serves as a style image. This allows for the artist to interact with a computational version of his past work, a key aspect computational creativity has to offer.

Also in Figure \ref{fig:iterative_same_portrait}, the photographic portrait is an input in the first canvas. The subsequent machine only uses as content or style image the preceding paintings. The observed divergence is hence an intertwined responsibility between the painter and the algorithms.

\section{Discussion}

Neural style transfer algorithms are computationally creative in the sense that they may produce new images with an aesthetic that can significantly differ from what a painter would do. 
In order to turn this creativity into artworks, we have specified various painting experiments on a real canvas between a painter and outputs from these algorithms.
We now report how these attempts shed light on a few aspects of the computational creativity of neural style transfer algorithms and cast them, in this specific setting, as \textit{computational catalysts} to human creativity.
Besides, the interactive painting process itself was designed to embody some questions related to computational creativity and to human-machine interplay, which has arguably become a major societal theme.

\paragraph{Computational Creativity and Catalyst.}
The initial motivation for designing human-neural-style-transfer interactive experiments was to create a single object out of many different style transfer outputs, focusing here on a painting instead of a printed version of the numerical output. This echoes other creative works with machines where some artists playfully worded themselves as \textit{editors} of the machine creativity, see for instance the rationale surrounding the last A.I. assisted musical album \textit{Chain Tripper} of the band \citet{yacht2019chain}.
Though during our painting experiments, the intertwining between the machine's outputs and painter interpretation was non-trivial since the painter was altering the machine suggestions. The painter felt the outputs were giving new style directions, wording them as \textit{computational catalysts} to his own creativity.

In these interactions with algorithms, we exploit the ability of style transfer methods to produce outputs based on the previous works of the painter. This is a simple yet powerful idea that allows an artist to interact with computationally influenced versions of its own (past) work. This was felt by the painter as a semi-extraneous interpretation of his past techniques, allowing him to rediscover some elements of his old practice in a surprising way. Besides our specific framework, this seems to be another major benefit and specificity of computational creativity.

Importantly also in these portrait paintings, the artist could not see, except in the beginning, the real photographic portrait. We purposely designed it in this way so that the painting practice could embody the fact of perceiving the world only through the machines' lens.
This has important societal echoes; for instance, the issues raised by the so-called \textit{fake news} stems from generative algorithms capacities, on a technical point of view, but, on the societal point of view, from our increasingly resorting to numerical pieces of information as a way to perceive the world. Here we hence implicitly explore what a painter felt when relying only on machine outputs to see the portraits.

Alternatively, it also gives another perspective on computationally creative algorithms, as offering new inspirational spaces to portray. Indeed we may not only explore algorithms outputs through printed versions, pretty much as we do not capture nature only through photography. Computationally creative outputs may hence be thought of as new types of landscapes for painters to capture.

Note also that the \textit{transient} essence of these \textit{computational landscapes} has very different rules than that of Nature; by erasing the content files or algorithms outputs, the painting could remain the only imprint of the machine outputs. This again is a specificity of computational creativity, when framed as a theme creator for artists, that is worth exploring.

\paragraph{Designing Human-Machine painting processes.}
A major aspect in these human-machine interactive processes is that we engineered the numerical outputs out in the real world, rather than having a painter to interact with machines on a numerical tablet for instance.

Indeed when the painting process materializes in a numerical tablet, it strongly constrains the painter's sensations; he does not feel the brushstrokes' gesture, the canvas is not perceived in the full-dimensional space, etc. Even with interactive experiments on a real-canvas, the painter felt some processes as being too intrusive or constraining, like the experiment reported in Figure \ref{fig:iterative_same_portrait} which forces the artist to follow unusual rules for creating. This highlights that computational creativity, when considered in such a human-machine interplay, is notably conditioned by the current state in the engineering of such interactive systems. For instance how much projection is less intrusive than a robotic arm? 

This level of machine's \textit{intrusion} is inherently linked with how the computational creativity of the algorithms is perceived, notably concerning the creative agency that is attributed to the machine outputs. Part of the discussion around computational creativity may hence be tightly related to some artists' feelings of losing a share of the creative agency when algorithms become more than a disposable tool.

So it appears that when engineering such systems, there are typically two directions in the interfacing, either the machine goes out of the numerical world or reversely, the human interacts with the machine in the numerical world. And as we described previously, the interfacing puts the human artists in very different situations. However, this may not only be considered as a limitation as each constraint forces the painter to embody what we may feel in our daily interaction with machines. In particular then, each type of interfacing echoes, and may advocate then, a different societal relation humans have with machines; at the era of machines, it is primal to explore many such experiments.

\paragraph{Plastic point of view.}
These interactive painting experiments were also designed to explore pictorial aspects. 

For instance the photographic portrait that initiate the series in Figure \ref{fig:thomas_portrait_multiple}-\ref{fig:iterative_same_portrait} were in black and white. However, the style transfer algorithm and the painter were not constrained in the grey-scale space. The painter could observe in projected outputs of the machine, or conversely initiate in the real canvas, the emergence of the colours. For instance, in Figure \ref{fig:thomas_portrait_multiple}, red appears on the eyebrows while in Figure \ref{fig:iterative_same_portrait} the colours are intended as variations of shade, which only exists through the machine.

While this is interesting from the creation point of view, it is also from the observer who is concerned about agency attribution. For a given aspect of the painting, like colours, did the painter simply repeat the machine colourization outputs, re-interpreted it or even started it? This reinforces the importance, in an exhibition, of algorithms' outputs as testimonies of the final artworks.

\section{Conclusion} We present interactive painting experiments between neural style transfer outputs and a painter.
It reveals many potential benefits of leveraging computational creativity in this type of interactive framework and questions some computation aspects of neural style transfer specifically.

\newpage

\subsubsection*{Acknowledgments}
The authors would like to thanks Tomas Angles for accidentally giving a new direction to our work; Fabienne Colin for her enthusiasm in exploring the machine style transfer for editing her own painting style, as well as pointing us to the series of Monet; Thibault S\'ejourn\'e for helpful discussion on Optimal Transport; Vivien Cabannes and John Zarka for rereading and Stephane Mallat for an interesting discussion. Finally Thomas would like to thanks Sebastian Pokutta for hosting him at the Zuse Institute in Berlin, where part of this work was carried out.

\bibliographystyle{iccc}
\bibliography{biblio}{}

\begin{thebibliography}{}

\bibitem[\protect\citeauthoryear{Cabannes \bgroup et al.\egroup
  }{2019}]{DialogCanvasMachine}
Cabannes, V.; Kerdreux, T.; Thiry, L.; Campana, T.; and Ferrandes, C.
\newblock 2019.
\newblock Dialog on a canvas with a machine.
\newblock In {\em NeuriIPS 2019 Workshop on Machine Learning for Creativity and
  Design}.

\bibitem[\protect\citeauthoryear{Cheng \bgroup et al.\egroup
  }{2019}]{cheng2019structure}
Cheng, M.-M.; Liu, X.-C.; Wang, J.; Lu, S.-P.; Lai, Y.-K.; and Rosin, P.~L.
\newblock 2019.
\newblock Structure-preserving neural style transfer.
\newblock {\em IEEE Transactions on Image Processing} 29:909--920.

\bibitem[\protect\citeauthoryear{Chung}{2015}]{chung2015drawing}
Chung, S.
\newblock 2015.
\newblock Drawing operations, 2015.

\bibitem[\protect\citeauthoryear{Cuturi}{2013}]{cuturi2013sinkhorn}
Cuturi, M.
\newblock 2013.
\newblock Sinkhorn distances: Lightspeed computation of optimal transport.
\newblock In {\em Advances in neural information processing systems},
  2292--2300.

\bibitem[\protect\citeauthoryear{Eitz, Hays, and Alexa}{2012}]{eitz2012hdhso}
Eitz, M.; Hays, J.; and Alexa, M.
\newblock 2012.
\newblock How do humans sketch objects?
\newblock {\em ACM Trans. Graph. (Proc. SIGGRAPH)} 31(4):44:1--44:10.

\bibitem[\protect\citeauthoryear{Ganin \bgroup et al.\egroup
  }{2018}]{ganin2018synthesizing}
Ganin, Y.; Kulkarni, T.; Babuschkin, I.; Eslami, S.; and Vinyals, O.
\newblock 2018.
\newblock Synthesizing programs for images using reinforced adversarial
  learning.
\newblock {\em arXiv preprint arXiv:1804.01118}.

\bibitem[\protect\citeauthoryear{Gatys \bgroup et al.\egroup
  }{2017}]{gatys2017controlling}
Gatys, L.~A.; Ecker, A.~S.; Bethge, M.; Hertzmann, A.; and Shechtman, E.
\newblock 2017.
\newblock Controlling perceptual factors in neural style transfer.
\newblock In {\em Proceedings of the IEEE Conference on Computer Vision and
  Pattern Recognition},  3985--3993.

\bibitem[\protect\citeauthoryear{Gatys, Ecker, and
  Bethge}{2015}]{gatys2015neural}
Gatys, L.~A.; Ecker, A.~S.; and Bethge, M.
\newblock 2015.
\newblock A neural algorithm of artistic style.
\newblock {\em arXiv preprint arXiv:1508.06576}.

\bibitem[\protect\citeauthoryear{Ghiasi \bgroup et al.\egroup
  }{2017}]{ghiasi2017exploring}
Ghiasi, G.; Lee, H.; Kudlur, M.; Dumoulin, V.; and Shlens, J.
\newblock 2017.
\newblock Exploring the structure of a real-time, arbitrary neural artistic
  stylization network.
\newblock {\em arXiv preprint arXiv:1705.06830}.

\bibitem[\protect\citeauthoryear{Goodfellow \bgroup et al.\egroup
  }{2014}]{goodfellow2014generative}
Goodfellow, I.; Pouget-Abadie, J.; Mirza, M.; Xu, B.; Warde-Farley, D.; Ozair,
  S.; Courville, A.; and Bengio, Y.
\newblock 2014.
\newblock Generative adversarial nets.
\newblock In {\em Advances in neural information processing systems},
  2672--2680.

\bibitem[\protect\citeauthoryear{Gupta \bgroup et al.\egroup
  }{2017}]{gupta2017characterizing}
Gupta, A.; Johnson, J.; Alahi, A.; and Fei-Fei, L.
\newblock 2017.
\newblock Characterizing and improving stability in neural style transfer.
\newblock In {\em Proceedings of the IEEE International Conference on Computer
  Vision},  4067--4076.

\bibitem[\protect\citeauthoryear{Huang and Belongie}{2017}]{huang2017arbitrary}
Huang, X., and Belongie, S.
\newblock 2017.
\newblock Arbitrary style transfer in real-time with adaptive instance
  normalization.
\newblock In {\em Proceedings of the IEEE International Conference on Computer
  Vision},  1501--1510.

\bibitem[\protect\citeauthoryear{Jing \bgroup et al.\egroup
  }{2017}]{jing17review}
Jing, Y.; Yang, Y.; Feng, Z.; Ye, J.; and Song, M.
\newblock 2017.
\newblock Neural style transfer: {A} review.
\newblock {\em CoRR} abs/1705.04058.

\bibitem[\protect\citeauthoryear{Johnson, Alahi, and
  Fei-Fei}{2016}]{johnson2016perceptual}
Johnson, J.; Alahi, A.; and Fei-Fei, L.
\newblock 2016.
\newblock Perceptual losses for real-time style transfer and super-resolution.
\newblock In {\em European conference on computer vision},  694--711.
\newblock Springer.

\bibitem[\protect\citeauthoryear{Jongejan \bgroup et al.\egroup
  }{2018}]{quickDraw}
Jongejan, J.; Rowley, H.; Kawashima, T.; Kim, J.; and Fox-Gieg, N.
\newblock 2018.
\newblock The quick, draw!-ai experiment.
\newblock {\em Mount View, CA, accessed Feb} 17.

\bibitem[\protect\citeauthoryear{Kleiner}{2009}]{livrecathedralerouen}
Kleiner, F.~S.
\newblock 2009.
\newblock Gardner's art through the ages: The western perspective, volume 2.

\bibitem[\protect\citeauthoryear{Kolkin, Salavon, and
  Shakhnarovich}{2019}]{TransferTransport}
Kolkin, N.; Salavon, J.; and Shakhnarovich, G.
\newblock 2019.
\newblock Style transfer by relaxed optimal transport and self-similarity.

\bibitem[\protect\citeauthoryear{Kotovenko \bgroup et al.\egroup
  }{2019}]{kotovenko2019content}
Kotovenko, D.; Sanakoyeu, A.; Ma, P.; Lang, S.; and Ommer, B.
\newblock 2019.
\newblock A content transformation block for image style transfer.
\newblock In {\em Proceedings of the IEEE Conference on Computer Vision and
  Pattern Recognition},  10032--10041.

\bibitem[\protect\citeauthoryear{Li and Wand}{2016}]{li2016precomputed}
Li, C., and Wand, M.
\newblock 2016.
\newblock Precomputed real-time texture synthesis with markovian generative
  adversarial networks.
\newblock In {\em European conference on computer vision},  702--716.
\newblock Springer.

\bibitem[\protect\citeauthoryear{Li \bgroup et al.\egroup
  }{2017}]{li2017universal}
Li, Y.; Fang, C.; Yang, J.; Wang, Z.; Lu, X.; and Yang, M.-H.
\newblock 2017.
\newblock Universal style transfer via feature transforms.
\newblock In {\em Advances in neural information processing systems},
  386--396.

\bibitem[\protect\citeauthoryear{Lu \bgroup et al.\egroup
  }{2017}]{lu2017decoder}
Lu, M.; Zhao, H.; Yao, A.; Xu, F.; Chen, Y.; and Zhang, L.
\newblock 2017.
\newblock Decoder network over lightweight reconstructed feature for fast
  semantic style transfer.
\newblock In {\em Proceedings of the IEEE International Conference on Computer
  Vision},  2469--2477.

\bibitem[\protect\citeauthoryear{Mroueh}{2019}]{WassersteinStyleTransfer}
Mroueh, Y.
\newblock 2019.
\newblock Wasserstein style transfer.

\bibitem[\protect\citeauthoryear{Nakano}{2019}]{nakano2019neural}
Nakano, R.
\newblock 2019.
\newblock Neural painters: A learned differentiable constraint for generating
  brushstroke paintings.
\newblock {\em arXiv preprint arXiv:1904.08410}.

\bibitem[\protect\citeauthoryear{Paszke \bgroup et al.\egroup
  }{2019}]{paszke2019pytorch}
Paszke, A.; Gross, S.; Massa, F.; Lerer, A.; Bradbury, J.; Chanan, G.; Killeen,
  T.; Lin, Z.; Gimelshein, N.; Antiga, L.; et~al.
\newblock 2019.
\newblock Pytorch: An imperative style, high-performance deep learning library.
\newblock In {\em Advances in Neural Information Processing Systems},
  8024--8035.

\bibitem[\protect\citeauthoryear{Risser, Wilmot, and
  Barnes}{2017}]{risser2017stable}
Risser, E.; Wilmot, P.; and Barnes, C.
\newblock 2017.
\newblock Stable and controllable neural texture synthesis and style transfer
  using histogram losses.
\newblock {\em arXiv preprint arXiv:1701.08893}.

\bibitem[\protect\citeauthoryear{Sanakoyeu \bgroup et al.\egroup
  }{2018}]{sanakoyeu2018style}
Sanakoyeu, A.; Kotovenko, D.; Lang, S.; and Ommer, B.
\newblock 2018.
\newblock A style-aware content loss for real-time hd style transfer.
\newblock In {\em Proceedings of the European Conference on Computer Vision
  (ECCV)},  698--714.

\bibitem[\protect\citeauthoryear{Schmitzer}{2019}]{schmitzer2019stabilized}
Schmitzer, B.
\newblock 2019.
\newblock Stabilized sparse scaling algorithms for entropy regularized
  transport problems.
\newblock {\em SIAM Journal on Scientific Computing} 41(3):A1443--A1481.

\bibitem[\protect\citeauthoryear{Szegedy \bgroup et al.\egroup
  }{2013}]{szegedy2013intriguing}
Szegedy, C.; Zaremba, W.; Sutskever, I.; Bruna, J.; Erhan, D.; Goodfellow, I.;
  and Fergus, R.
\newblock 2013.
\newblock Intriguing properties of neural networks.
\newblock {\em arXiv preprint arXiv:1312.6199}.

\bibitem[\protect\citeauthoryear{Ulyanov}{2016}]{ulyanov2016texture}
Ulyanov, D.
\newblock 2016.
\newblock Texture networks: Feed-forward synthesis of textures and stylized
  images.
\newblock Association for Computing Machinery.

\bibitem[\protect\citeauthoryear{Xie, Hachiya, and
  Sugiyama}{2013}]{xie2013artist}
Xie, N.; Hachiya, H.; and Sugiyama, M.
\newblock 2013.
\newblock Artist agent: A reinforcement learning approach to automatic stroke
  generation in oriental ink painting.
\newblock {\em IEICE TRANSACTIONS on Information and Systems} 96(5):1134--1144.

\bibitem[\protect\citeauthoryear{YACHT}{2019}]{yacht2019chain}
YACHT.
\newblock 2019.
\newblock Chain tripping.

\bibitem[\protect\citeauthoryear{Zhang \bgroup et al.\egroup
  }{2018}]{zhang2018unreasonable}
Zhang, R.; Isola, P.; Efros, A.~A.; Shechtman, E.; and Wang, O.
\newblock 2018.
\newblock The unreasonable effectiveness of deep features as a perceptual
  metric.
\newblock In {\em Proceedings of the IEEE Conference on Computer Vision and
  Pattern Recognition},  586--595.

\bibitem[\protect\citeauthoryear{Zhang \bgroup et al.\egroup
  }{2019a}]{zhang2019multimodal}
Zhang, Y.; Fang, C.; Wang, Y.; Wang, Z.; Lin, Z.; Fu, Y.; and Yang, J.
\newblock 2019a.
\newblock Multimodal style transfer via graph cuts.
\newblock In {\em Proceedings of the IEEE International Conference on Computer
  Vision},  5943--5951.

\bibitem[\protect\citeauthoryear{Zhang \bgroup et al.\egroup
  }{2019b}]{GraphCut19}
Zhang, Y.; Fang, C.; Wang, Y.; Wang, Z.; Lin, Z.; Fu, Y.; and Yang, J.
\newblock 2019b.
\newblock Multimodal style transfer via graph cuts.

\end{thebibliography}

\end{document}